\documentclass[a4paper,UKenglish,cleveref, autoref, thm-restate]{lipics-v2021}

\usepackage{hyperref}
\usepackage{xspace}

\usepackage[T1]{fontenc}
\usepackage{graphicx, amsmath,bm}
\usepackage{colortbl}
\usepackage{hhline}
\usepackage[title]{appendix}
\usepackage{etoolbox}
\usepackage{xcolor}

\usepackage{tikz, pifont}
\usepackage{stackengine}
\usepackage{subcaption}

\usepackage{makecell}
\usepackage{array}
\renewcommand{\qed}{\hfill\blacksquare}
\usepackage{setspace}

\definecolor{YellowGreen}{rgb}{0.75, 0.9, 0.36}

\usepackage{multirow}
\newif\ifshowcomment
\showcommenttrue
\ifshowcomment
    \newcommand{\dawn}[1]{\textsf{\color{red}{[{Dawn: #1}]}}}
    \newcommand{\yujin}[1]{\textsf{\color{blue}{[{Yujin: #1}]}}}
    \newcommand{\korn}[1]{\textsf{\color{magenta}{[{Kornrapat: #1}]}}} 
    \newcommand{\ag}[1]{\textsf{\color{red}{[{AG: #1}]}}} 
    \newcommand{\ariah}[1]{\textsf{\color{red}{[{Ariah: #1}]}}}
    \newcommand{\philipp}[1]{\textsf{\color{teal}{[{Philipp: #1}]}}}
    \newcommand{\todo}[1]{\textsf{\color{orange}{[{TO DO: #1}]}}}
\else
    \newcommand{\dawn}[1]{}
    \newcommand{\yujin}[1]{}
    \newcommand{\korn}[1]{}
    \newcommand{\ag}[1]{}
    \newcommand{\ariah}[1]{}
    \newcommand{\philipp}[1]{}
    \newcommand{\todo}[1]{}
\fi

\title{What Drives the (In)stability of a Stablecoin?}

\author{Yujin Kwon}{UC Berkeley, USA}{yujinyujin9393@berkeley.edu}{ORCID}{}
\author{Kornrapat Pongmala}{UC Berkeley, USA}{kornrapatp@berkeley.edu}{ORCID}{}
\author{Kaihua Qin}{Imperial College London, UK}{kaihua.qin@imperial.ac.uk}{ORCID}{}
\author{Ariah Klages-Mundt}{Cornell, USA}{aak228@cornell.edu}{ORCID}{}
\author{Philipp Jovanovic}{UCL, UK}{p.jovanovic@ucl.ac.uk}{ORCID}{}
\author{Christine Parlour}{UC Berkeley, USA}{parlour@berkeley.edu}{ORCID}{}
\author{Arthur Gervais}{UCL, UK}{arthur@gervais.cc}{ORCID}{}
\author{Dawn Song}{UC Berkeley, USA}{dawnsong@cs.berkeley.edu}{ORCID}{}

\authorrunning{Kwon et al.} 

\Copyright{Jane Open Access and Joan R. Public} 



\nolinenumbers 

\begin{document}

\maketitle

\begin{abstract}
In May 2022, an apparent speculative attack, followed by market panic, led to the precipitous downfall of UST, one of the most popular stablecoins at that time. 
However, UST is not the only stablecoin to have been depegged in the past. Designing resilient and long-term stable coins, therefore, appears to present a hard challenge.

To further scrutinize existing stablecoin designs and ultimately lead to more robust systems, we need to understand where volatility emerges.
Our work provides a game-theoretical model aiming to help identify why stablecoins suffer from a depeg.
This game-theoretical model reveals that stablecoins have different price equilibria depending on the coin's architecture and mechanism to minimize volatility.
Moreover, our theory is supported by extensive empirical data, spanning $1$ year. To that end, we collect daily prices for 22 stablecoins and on-chain data from five blockchains including the Ethereum and the Terra blockchain.
\end{abstract}

\section{Introduction}
\label{sec:intro}

Cryptocurrencies, such as Bitcoin and Ethereum, have gained tremendous global attention over the past few years. 
Despite their popularity, the significant price volatility of their native coins, such as BTC and ETH, has hampered their usage as a consistent value storage medium. 
To address this volatility issue, \emph{stablecoins} were proposed and, since their introduction, have taken up a central role in the Decentralized Finance (DeFi) ecosystem. At the time of writing, stablecoins have reached a market capitalization of over USD \$116 billion. 

The goal of a stablecoin is to offer a store of value with low volatility, i.e., the fluctuation of the price of a stablecoin should be minimized. Since the price of an asset is always expressed relatively to the value of another asset, one popular method to define a coin with a ``stable value'', is to link or \emph{peg} a stablecoin to the value of a government-issued fiat currency such as the US Dollar. Note that pegged stablecoins aim to remain stable relative to the target currency, but may fluctuate with respect to other assets. The widespread adoption of fiat-pegged stablecoins crafts a productive connection between traditional and decentralized finance ecosystems.

Stablecoins can achieve price stability by adopting a pegging mechanism similar to a traditional pegged currency. For national currencies to achieve fixed exchange rates, the central bank plays an important role. The central bank adjusts the market supply and demand to attain the target exchange rate by holding foreign reserves that can be exchanged for national currency. For example, if an exchange rate is less than the target, the central bank purchases national currency from the market with foreign reserves, which increases the exchange rate~\cite{lyons2023keeps}. 

However, the blockchain use of a stablecoin diversifies a stablecoin design, allowing for a more progressive, albeit often complicated, pegging mechanism. For example, they can rely on cryptocurrencies to maintain the peg, and some stablecoins even attempt to stabilize a price without relying on reserves consisting of exogenous assets.\footnote{It refers to assets that are run independently of the stablecoin system. For example, Bitcoin and Ethereum are one of the most representative exogenous crypto assets.} 

In broad strokes, we distinguish between \emph{collateralized} and \emph{algorithmic} stablecoins, while hybrid models may exist. Collateralized stablecoins safeguard their value by holding a reserve of exogenous assets that can be used to purchase stablecoins from users. 
On the other hand, algorithmic stablecoins aim to achieve a peg through algorithmically expanding and contracting supply without holding exogenous collateral to back their coins. The systems mint an endogenous cryptocurrency to purchase stablecoins from users.

The most intuitive design of the pegging mechanisms of stablecoins is to allow users to buy/sell stablecoins from/to the system at near the target price (e.g., 1 USD) to maintain the peg. If the market price deviates from the target price, users have a monetary incentive to perform arbitrage between the stablecoin system and the secondary market. Depending on the types of assets stored in the reserves, we can categorize collateralized stablecoins into \textit{fiat-collateralized} or \textit{crypto-collateralized}.
\emph{Algorithmic} stablecoins also adopt such a pegging mechanism design, but they use endogenous cryptocurrencies. 

Moreover, stablecoins can be designed with close ties to lending systems, and such stablecoins are referred to as \emph{over-collateralized}. In these coin systems, users deposit collateral of a higher value than the stablecoins issued to them, thereby becoming debtors. When users redeem the stablecoins (i.e., debts) to the system, they can get back the deposited collateral. If a position is at risk of turning into a bad debt due to a drop in the collateral value, the system allows other users to liquidate the position by paying back the stablecoins on behalf of the debtors and thereby getting the underlying collateral~\cite{qin2021empirical}. In this design, if the market price is less than the target price, users are expected to buy stablecoins from the market to settle (their own or others') debts at a discount, which raises the overall market price to the peg.

In fact, a crypto-collateralized stablecoin is a term often associated with a broad categorization of any stablecoin backed by cryptocurrencies. However, in this study, we further partition this broad categorization based on how redemption takes place.
In other words, we differentiate between crypto-collateralized and over-collateralized stablecoins while both use cryptocurrencies as collateral; crypto-collateralized stablecoins purchase and sell stablecoins to users at a target price to maintain the peg. On the other hand, over-collateralized systems apply a lending system.


\begin{figure}
    \centering
    \includegraphics[width=0.7\columnwidth]{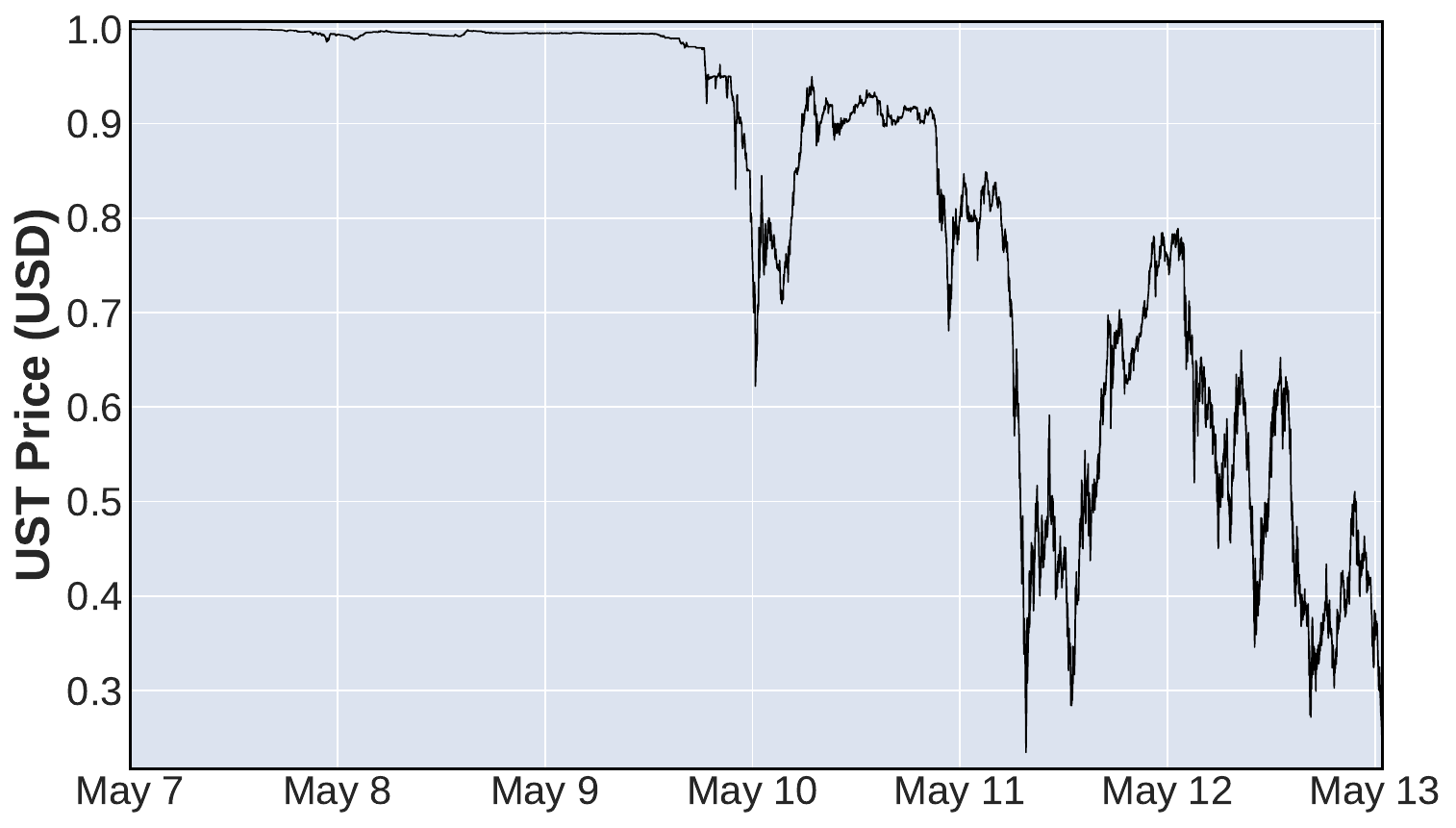}
    \caption{UST spot price (UST/USDT market) on Binance (May~7 -- May~13, 2022).}
    \label{fig:binance-ust-usdt-price}
\end{figure}

Although stablecoins are designed for relative price stability, price history shows that it is quite common for pegs to break. The Terra stablecoin, UST, which was one of the most representative algorithmic stablecoins, collapsed in the second week of May 2022 (see Figure~\ref{fig:binance-ust-usdt-price})~\cite{terra1,terra2}. 
As an algorithmic stablecoin, Terra uses an endogenous cryptocurrency, Luna, to maintain the peg of UST at 1 USD; the system mints 1 USD worth of Luna to the users for each UST they redeem to the system. 
Therefore, if the market price of UST falls below 1 USD, users should sell their coins to the system instead of the market for a higher profit. 
However, this redemption process has a risk; it inflates the market with the minted Luna, driving the price of Luna down, scaring users to sell their coins, and making each subsequent UST redemption mint even more Luna. 
As a result, a significant deviation from the target price could lead to a larger deviation.
This is commonly known as the \emph{death spiral}, which is what happened to Terra in May 2022. Another example of a USD-pegged stablecoin is Tether, where there have been doubts regarding the reserve coverage, raising concerns about bank run risks~\cite{klages2020stablecoins,arner2020stablecoins}.
Tether was even fined by the US Commodity Futures Trading Commission (CFTC) for making ``untrue or misleading statements and omissions'' regarding their reserves~\cite{tether_law,tether_fine}.
In short, many stablecoin designs contain weaknesses that may threaten price stability, and in some cases these have led to the collapse of their prices to almost zero~\cite{terra1,terra2,terra3,terra4,nubits}.

\subsection{Implications of our work}

The statistically observed difference in price volatility of stablecoins~\cite{jarno2021does} and several collapse events including Terra UST beg the question of how mechanism design choices get translated into stablecoin price stability. 
Currently, it is unclear how and to what extent these design-dependent risks are reflected in the actual stablecoin price stability. 
Bearing the risks does not necessarily directly lead to actual price instability. 
For example, although many people point out a bank run risk\footnote{It can be triggered by a lack of reserves stored by a stablecoin system.} of Tether, we also often observe that it did not significantly affect its price in the real world. Even when Tether was fined multiple times due to some evidence that may suggest holding partial reserves, its price was still around \$1. 
Indeed, in general, fiat-collateralized stablecoins that store reserves in fiat money are regarded as more stable assets than other stablecoin types~\cite{jarno2021does}.
In contrast, many crypto-collateralized stablecoins that hold collateral in cryptocurrencies often depreciate below \$1 and return more slowly to the pegging state. 
In particular, considering when UST depegged, unlike the Tether case at that time, UST could not quickly return to the peg even though the market cap of Luna, which is the asset backing it, exceeded its market cap in the early stage of the catastrophic event~\cite{terra1,terra2,terra3,terra4}. 
These suggest that the relationship between design risks and price instability is not straightforward, which makes it hard to understand their inherent relationship clearly. 

The goal of this work is to assist in developing a solid, common theory that can model the relationship between stablecoin designs and price volatility and further compare the effects of their design differences. 
Our work helps to understand why stablecoins show different price volatility by modeling them in a common framework, taking into account the difference in asset types that various stablecoin systems use to purchase coins from users. The type of asset paid to users by a system can matter; a volatile asset can lower the user's recognized payment value due to a price fluctuation during the payment transaction process. We show that this factor would significantly affect a level of stablecoin price stability. In practice, when designing a stablecoin, developers can focus more on addressing risks related to a lack of reserves while little considering the discrepancy in payment values between users and the system~\cite{usdn,lusd,celo,kereiakes2019terra}. This suggests that the current stablecoin systems may be missing the critical point. 

\subsection{Related work}

Many works~\cite{klages2020stablecoins,arner2020stablecoins,kwon2021trilemma,moin2020sok,algorithmic} have analyzed the systematic risks of different stablecoin designs. 
As discussed in \cite{klages2020stablecoins}, traditional models of bank runs \cite{diamond1983bank}, runs on currency pegs \cite{morris1998unique}, and pegged redemption money market funds \cite{parlatore2016fragility} can be applied to understand the risks of various stablecoins; it is possible to reinterpret the pegging mechanism as the central bank, and so basic results about bank and currency runs can translate to many stablecoin settings.
Given this, \cite{routledge2022currency} investigates a way to avoid speculative attacks~\cite{morris1998unique} in partially backed stablecoins, and shows that optimal exchange rate policies rate can help avoid the risk. 
\cite{li2022money} also studies the optimal strategy of collateralized stablecoin systems to investigate whether stablecoins are sustainable given the risks. 



On the other hand, models explaining the price (in)stability of stablecoins are relatively sparse. Note that having risk in a stablecoin design does not always lead to price drops (e.g., the Tether case), and thus studying systemic risks alone may not be sufficient to understand stablecoins. \cite{klages2022while,klages2021stability} model stablecoins like DAI, where issuance is based on a market for leverage. In this context, they characterize deleveraging spirals that caused instability in DAI on ``Black Thursday'' in March 2020. 
\cite{lyons2023keeps} models how stablecoins backed by 100\% reserves, such as USDC, maintain price stability through arbitrage with minting and redemption. \cite{klagesmundt2022designing} explains a shape of redemption curves of several stablecoins and then designs a redemption curve that attains price stability.

Beyond few attempts to elucidate the price (in)stability of stablecoins, existing models are difficult to generalize due to the heterogeneity of the stablecoin design space.
Moreover, various distinctions about stablecoin models are not as developed, such as modeling the fact that reserve assets are not the currency target (e.g., many stablecoins hold reserves consisting of cryptocurrencies, not their price target). 
We address this by proposing a game-theoretical model motivated by \cite{morris1998unique} as a starting point. We then extend to include reserve assets that change in price and further by breaking down assets backing into exogenous and endogenous assets, which have different degrees of price volatility (this is encoded in the function $r^c$ in our model).

Having a common framework to analyze stablecoin price stability would be conducive not only to understanding the connection between stablecoin designs and price stability but also to comparing different stablecoin designs.
In addition, there is room to expand a common model to better characterize nuanced differences across different types of stablecoins.
This is the approach we take in encoding a model that is flexible enough to analyze redemption mechanisms across different types of stablecoins.


\smallskip\noindent\textbf{Contributions.}
Here, we summarize the contributions of this paper below.

\begin{itemize}
    \item We build a game theoretical model to analyze and compare four different stablecoin designs, fiat-collateralized, crypto-collateralized, algorithmic, and over-collateralized stablecoins, under a general framework. 
    \item We analyze the game model for each stablecoin design and show that they result in different price equilibria, allowing us to relatively rank the price stability of stablecoin designs. 
    Specifically, we show that fully backed fiat-collateralized stablecoins can guarantee the peg by proving they have the unique pegging equilibrium for any economic condition. 
    On the other hand, our analysis shows that partially backed fiat-collateralized stablecoins have multiple equilibria including the pegging state; in particular, the pegging state becomes a self-fulfilling equilibrium. 
    The equilibria of crypto-collateralized, algorithmic, and over-collateralized stablecoins depend on economic situations. We show that crypto-collateralized and algorithmic stablecoins have the unique pegging equilibrium in a good economic environment, but there are only depegging equilibria in poor economic conditions. Even where they can hold sufficient reserves to back the stablecoins fully, they may have only depegging equilibria. 
    Lastly, while analyzing the equilibria of over-collateralized stablecoins, we find that they have multiple equilibria including the pegging state in most economic conditions.
    In other words, it may be difficult to rank crypto-collateralized, algorithmic, and over-collateralized stablecoins absolutely; in a good economic condition, crypto-collateralized and algorithmic stablecoins are better than over-collateralized stablecoins because crypto-collateralized and algorithmic stablecoins have the unique pegging equilibrium. But over-collateralized stablecoins can be superior to the two in poor economic conditions. 
    \item We collect and analyze the daily price of stablecoins and redemption transactions to empirically confirm our theories. We compare price stability by stablecoin type and find a statistical connection between a stablecoin price and the payment that users get when redeeming stablecoins.
    As a result, we find that a stablecoin price has overall fluctuated in agreement with our theory. 
\end{itemize}






\smallskip\noindent
\textbf{Paper structure:}
The rest of the paper is organized as follows. Section~\ref{sec:model} proposes a general, common game-theoretical model to analyze different stablecoin designs. Section~\ref{sec:type} represents four types of stablecoins under the proposed game-theoretical framework. Then, we analyze the game and find equilibria in Section~\ref{sec:eq}. Section~\ref{sec:data} provides empirical analyses to support our theories. Lastly, we discuss stablecoin design challenges in Section~\ref{sec:discuss} and conclude in Section~\ref{sec:conclusion}.
\section{Model}
\label{sec:model}

In this section, we provide a possible avenue to model a stablecoin system using a game-theoretical framework. Based on the stablecoin model, we will analyze four types of stablecoins and quantify their price (in)stability degrees later. We first present Table~\ref{tab:par} that summarizes the parameters used throughout the paper.

\begin{table*}[!h]
    \centering
    \begin{tabular}{>{\centering\arraybackslash}m{0.2\textwidth}|>{\raggedright\arraybackslash}m{0.8\textwidth}}
        Notation & \multicolumn{1}{c}{Definition}  \\
        \hline
        \hline
        $\theta$ & A state of fundamentals \\
        \hline
        $e(\theta)$ & A reached stablecoin price for given economic state $\theta$ without any system intervention\\
        \hline
        $M$ (or $M^\prime$) & The total quantity of coins that users choose to sell to the market at the moment (or in the future) \\
        \hline
        $p(M)$ (or $P(M^\prime)$) & A stablecoin price for given $M$ (or $M^\prime$) \\
        \hline
        $v$  (or $v^\prime$) & A value that the system pays users who redeem their stablecoins at the moment (or in the future)\\
        \hline
        $i(\cdot)$ & An incentive function for users to keep holding their stablecoins\\
        \hline
        $Q$ & The total quantity of coins whose users want to redeem at one point\\
        \hline
        $V^f$ & The total value of fiat reserves in a fiat-collateralized stablecoin\\
        \hline
        $c$ & A cryptocurrency used to back stablecoins in crypto-collateralized, algorithmic, and over-collateralized systems\\
        \hline
        $p^c_u$ & $c$'s price to which users refer\\
        \hline
        $p^c_s$ & $c$'s price to which the system assumes \\
        \hline
        $V^c(\theta)$ & A total value of crypto reserves in crypto-collateralized stablecoins\\
        \hline
        $r^c\left(Q, \theta\right)$ & A ratio between $p^c_u$ and $p^c_s$ \\
        \hline
        $D^L(\theta)$ & The total quantity of stablecoins that users should redeem in the liquidation process of over-collateralized systems at one point\\
        \hline
        $D_{u_i}(\theta)$ & The stablecoin debt of user $u_i$ that did not enter the liquidation process in over-collateralized systems \\
        \hline
        $o(\theta)$ & The system's estimated value of cryptocurrencies that an over-collateralized system pays to users who redeem their stablecoins  \\
        \hline
    \end{tabular}
    \caption{List of parameters}
    \label{tab:par}
\end{table*}

\subsection{Game-theoretical framework}
\label{subsec:system}

If a stablecoin is to be pegged at 1, the pegging state is defined as $p^s=1$, where $p^s$ indicates the current market price of the stablecoin. 
A pegging mechanism of the stablecoin system intervenes in the exchange market to maintain the peg. 
Considering that many stablecoins are currently suffering from downward price instability rather than upward price instability, in this paper, we focus on the mechanism that recovers a price when a stablecoin depreciates below 1. 
If the price falls below 1, the mechanism tries to decrease market supply and increase market demand by incentivizing users not to sell and to buy stablecoins in the market. 
Thus, designing a proper payoff function of users would be a key to achieving price stability. 
Note that we will not consider that the stablecoin system can halt the exchange market like a circuit breaker. 

If a pegging mechanism functions effectively, $p^s$ will eventually reach its target value of 1. However, in the absence of any system intervention in the market to adjust the coin price (i.e., if a pegging mechanism gives up a price recovery), its equilibrium price is determined by the economic characteristics, referred to as the fundamental state $\theta$~\cite{morris1998unique}. A higher value of $\theta$ signifies ``stronger fundamentals''. 
In other words, a high $\theta$ indicates a favorable economic condition where assets can appreciate. 
We express the coin price without system intervention as an increasing function $e(\theta)$ of $\theta$, where the value of $e(\theta)$ is assumed to be always below 1. 

\begin{figure}[h]
    \centering
    \includegraphics[width=0.6\columnwidth]{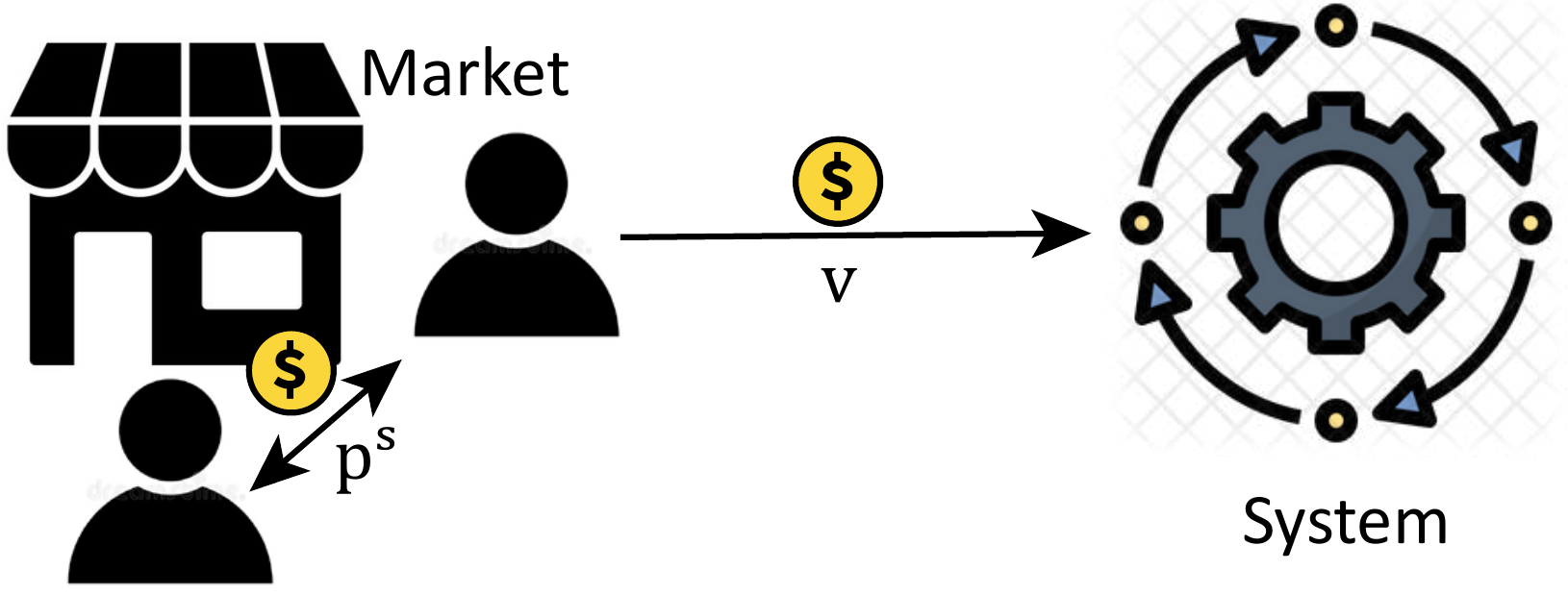}
    \caption{Relationships between users, the market, and the system. In the market, users can trade their stablecoins with each other at the current market price $p^s.$ Also, users can redeem their stablecoins to the system, where they sell their stablecoins to the system and receive $v.$}
    \label{fig:model}
\end{figure}

In this paper, we model a one-shot game with many users. Note that, currently, numerous users trade stablecoins in the market. 
Users can transact their coins with each other in the exchange market and issue coins from or redeem them to the system according to the pegging mechanism. 
Because we focus on the mechanism that recovers a price below $1$, it is enough to consider only redemption as an interaction between users and the system. Figure~\ref{fig:model} shows the relationship between users, the market, and the system. 

On the supply side, stablecoin holders can select their actions among 1) selling their coins in the market, 2) redeeming coins to the system, and 3) holding coins continuously. 
If they decide to sell coins to the market, their payoff would be the current stablecoin price $p^s.$ If they choose to redeem coins to the system, they are paid a value $v$ by the system, where $v$ is characterized according to each stablecoin design.
Lastly, when they keep holding their coins, their payoff depends on the future value of the stablecoin and an incentive provided by the system for users to keep holding their stablecoins (e.g., savings interest for a stablecoin). 
Here, let $i(\cdot)$ denote an incentive function for users to keep holding stablecoins, and it has a stablecoin value as input of a function.\footnote{For simplicity, we will omit a discount factor or time discounting because it does not significantly affect the results.}
The future stablecoin value can be expressed as $\max\{p^{s\prime},v^\prime\}$, where $p^{s\prime}$ indicates the price at which users will trade coins in the market in the future and $v^\prime$ indicates the redemption value that users can get from the system when redeeming coins in the future.\footnote{Note that we omitted a time variable for simplicity.}  
As a result, the payoff of users who decided to keep holding their stablecoins can be expressed as $\max\{i(p^{s\prime}),i(v^\prime)\}$ (or it is also possible to be expressed as $i\left(\max\{p^{s\prime},v^\prime\right)\})$).\footnote{In fact, users cannot be sure about the future value of the stablecoin. Therefore, in a more advanced model setting, the expected payoff would be expressed as $\int_{x} i(x)f\left(\max\{p^{s\prime},v^\prime\}=x\right)dx$ (i.e., an expected value based on the user's expectation on the future stablecoin value), where $f(\max\{p^{s\prime},v^\prime\})$ is a probability density function based on user expectation on $\max\{p^{s\prime},v^\prime\}.$ However, in this work, we simplify it, assuming users believe a specific value of the future stablecoin price for sure, because the simplification does not change the results and implications.} 
Note that $i(x)$ is equal to $x$ when the system provides no incentive for users to keep their coins (e.g., zero savings interest). Meanwhile, the greater the incentives, the greater the value of the function $i$.

In terms of the demand side, potential buyers can select their actions among 1) not buying coins in the market (i.e., not joining the stablecoin market), 2) buying coins in the market and immediately redeeming the coins to the system, and 3) buying coins in the market and keeping the coins in their pocket. Note that their actions are done in the one-shot game. 
In fact, their payoff equals the above described payoff (of stablecoin holders) minus $p^s$. Specifically, if they decide not to buy coins, their payoff would be 0. On the other hand, if they decide to go for the second action, their payoff would be $v-p^s$. Lastly, if they choose the third action, their payoff would be $\max\{i(p^{s\prime}),i(v^\prime)\}-p^s$.
Given this, the game analysis would be the same with considering only the payoff of stablecoin holders on the supply side. Therefore, for simplicity, we will consider only the payoff of stablecoin holders. 

According to the basic economic theory, the stablecoin price is determined by the market supply and demand. In our model, the stablecoin price is expressed as a function of only the market supply from the above paragraph. The market supply is related to the set of all users who decide to sell their coins to the market. When $M$ denotes the total quantity of coins that the users decide to sell to the market at the moment, $p^s$ can be expressed as $p(M),$ where $p$ indicates a decreasing function that converts from $M$ to a price. 
Therefore, $p(M)$ would decrease and increase if more users choose to sell their coins in the market (i.e., an increase in $M$) and choose other actions (i.e., a decrease in $M$), respectively. 
Similarly, we will express the future stablecoin price $p^{s\prime}$ as $p(M^\prime)$, denoting the future market supply by $M^\prime.$
Lastly, we assume that $p(\cdot)$ is always equal to or less than 1 to focus on when a stablecoin depreciates below 1.

In summary, in our model, the payoff function of users is as follows. 

\begin{equation}
\text{payoff}=\begin{cases}
p(M) (=p^s) &\text{if they sell coins to the market,}\\
\vspace{2mm}
v &\text{if they redeem coins to the system,}\\
\vspace{2mm}
\max\{i(p(M^\prime)),i(v^\prime)\}  &\parbox[m]{0.5\linewidth}{if they keep holding coins}
\end{cases}  
\label{eq:pay}
\end{equation}

\subsection{A unique pegging equilibrium}
\label{subsec:unique}

A stablecoin system should achieve $p(M)=1.$ Adopting $v$ and $v^\prime$ appropriately is essential to stabilize the coin price because it affects the decision of rational users. 
To guarantee the peg, there should be a reachable, unique equilibrium in which $p(M)=1.$ 
Here, that the state $p(M)=x$ is an \textit{equilibrium} indicates that a user's rational decision on whether to keep or change its action to increase their expected payoff in that state cannot change the stablecoin price $p(M)$ (i.e., the stablecoin price is fixed at $x$). 
Moreover, we say the state is \textit{reachable} if rational actions of users make the stablecoin price $p(M)$ return to $x$ in the case where $p(M)\not=x$; for example, if $p(M)$ is less than $x,$ rational users sell fewer coins to the market, which would recover $p(M)$ to $x$ by decreasing a value of $M.$ 

A sound stablecoin system should choose a proper value of $v$ and $v^\prime$ that equilibrates only the state $p(M)=1.$ 
The following theorem presents two sufficient conditions and one necessary condition to make the pegging state (i.e., $p(M)=1$) the reachable and unique equilibrium.

\begin{theorem}
\label{thm:unique}
To have a reachable and unique equilibrium as $p(M)=1,$ the following two conditions are sufficient:
$$\max\{v,i(v^\prime),i(p(M^\prime))\}>p(M) \text{ if } p(M)<1,$$ and 
$$\max\{v,i(v^\prime),i(p(M^\prime))\}\geq 1 \text{ if } p(M)=1.$$
Moreover, the first condition is necessary. 
\end{theorem}

The theorem says that to ensure a unique pegging equilibrium, we must design a stablecoin as follows: having a sufficiently high $v$ or $v^\prime$, or designing a (large) incentive function $i(\cdot)$ for users to keep holding their coins. 

Consider when there is little incentive (i.e., $i(x)\approx x$). Then Theorem~\ref{thm:unique} implies that, to guarantee the peg, either $v$, $v^\prime$, or $p(M^\prime)$ must be greater than $p(M)$ for any $p(M) <1.$ 
In the case where the market price of the stablecoin is less than the target price, if $v$ or $v^\prime$ is high enough to satisfy this, users would redeem their coins to the system or keep holding coins, instead of selling them now in the market, which decreases $M$ and increases the market price. 
On the one hand, if users expect the stablecoin price to increase (i.e., $p(M^\prime)>p(M)$) due to the promising economic situation, they would keep holding coins rather than selling them immediately in the market, which also naturally increases the market price. However, recall that we assumed that the stablecoin price without any system intervention in the market is $e(\theta)$ ($<1$) in Section~\ref{subsec:system}. 
Therefore, in this paper, we do not consider the situation where the price will naturally increase to 1 without any pegging mechanism by an economic uptrend. 

However, this is not the only way to ensure a reachable and unique equilibrium as the state that $p(M)=1.$
If a system can give users sufficient incentives to hold coins, it is also possible to, even with low $v$, $v^\prime$, and $p(M^\prime)$, make the pegging state the unique equilibrium.
Note that with a high value of the function $i$, $i(v^\prime)$ and $i(p(M^\prime))$ can be greater than $p(M).$ 
As great incentives, we can come up with large savings interest for a stablecoin. Intuitively, such incentives are conducive to a decrease in market supply necessary for a price increase.
We present the proof of Theorem~\ref{thm:unique} in Appendix~\ref{app:proof1}.

\section{Representing each stablecoin type with the model}
\label{sec:type}


In this section, we describe how $v$ is characterized in various types of stablecoins. Here, $v$ and $v^\prime$ would be the same because all stablecoins analyzed in this paper have a pegging mechanism independent of time. 
First, to ease the reader's understanding, we start with an intuition justifying how we will represent each type of stablecoins with our model. While a stablecoin system pays users (more than) the target value (e.g., \$1) upon redemption to prevent its price from falling below the target value, the payment that users realize can be lower than the intended value, depending on the type of assets paid. 

For example, consider a stablecoin system that holds cryptocurrencies as reserves to maintain the peg. It is widely known that cryptocurrencies experience high price volatility, which can result in unforeseen stability-threatening events in these stablecoin systems. Even if the systems were to reward a user with \$1 in cryptocurrency, the user might not end up receiving assets worth \$1. That is because there exist several challenges, such as delayed price oracles, as well as the sheer time difference between transaction creation and execution. 
The challenges cause payment value discrepancies between users and the system. 
Indeed, we have experienced stablecoins depegging permanently because their oracle update interval was set to 10 minutes (cf. Iron~\cite{Analysis10:online}).
As a result, the stablecoin's price stability can be threatened if the collateral is a volatile asset. 

The situation deteriorates if a system uses assets with higher risk (e.g., endogenous cryptocurrencies) to back its stablecoin.
Such assets would receive greater downward market pressure from the stablecoin redemption process in which the system should release the assets on the market.
The downward pressure can lead to liquidation spirals and market panics of the reserve assets, leading to a significant price drop and swiftly pushing a stablecoin toward an unhealthy state. A famous example of such a downfall is the severe price drop of Luna (an endogenous asset in the UST system) due to significant UST redemptions by users in May 2022. 
In this case, even though the Luna system paid users \$1 based on its Luna price oracle, users did not expect to earn \$1 due to a rapid decline in Luna's price.

With these intuitions in mind, we characterize $v$ of various types of stablecoins. Table~\ref{tab:v} summarizes $v$ of stablecoins that will be described below. 

\begin{table}[!ht]
    \centering
    \begin{tabular}{c|>{\raggedright\arraybackslash}m{0.8\columnwidth}}
        Type & \multicolumn{1}{c}{$v$} \\
        \hline
        \hline
        Fiat & $1\,\,$ if $Q\leq V^f,\,\,$ otherwise $V^f/Q$ \\
        \hline
        Crypto & $r^c\left(Q, \theta\right)\,\,$ if $Q\leq V^c(\theta),\,\,$ otherwise $r^c\left(Q, \theta\right)\cdot V^c(\theta)/Q$ \\
        \hline
        Algo & $r^c\left(Q, \theta\right)$ \\
        \hline
        Over & $r^c\left(Q, \theta\right)\cdot o(\theta)\,\,$ if $0< D^L(\theta)$ or $0< D_{u_i}(\theta),\,\,$ otherwise 0  \\
        \hline
    \end{tabular}
    \caption{$v$ by stablecoin type}
    \label{tab:v}
\end{table}

\subsection{Fiat-collateralized stablecoins}

We first look at fiat-collateralized stablecoins such as USDT and USDC. 
In these stablecoins, the system issues coins and purchases them from the market at 1 in fiat currency to achieve the peg. 
However, taking coins out of the market is possible only when the system's fiat reserves are not exhausted. 
We let $Q$ and $V^f$ denote the total quantity of coins whose holders want to redeem at one point and the total value of the fiat reserves at the moment, respectively. 
If $Q\leq V^f$, users can always be paid 1, and $v$ is 1. 
On the other hand, if $Q> V^f,$ users have two cases where they receive 1 or not. Then, under the assumption that users are uniform, the expected value would be $V^f/Q$ because the probabilities of users to get 1 and 0 is $V^f/Q$ and $1-V^f/Q$, respectively. Therefore, in that case, the expected value of $v$ would be $V^f/Q$. 

\subsection{Crypto-collateralized stablecoins} 

These stablecoins\footnote{Note that they are different from over-collateralized stablecoins in this paper.} (e.g., USDN) are similar to fiat-collateralized stablecoins except that the reserves are stored in cryptocurrencies. 
Specifically, the pegging mechanism pays 1 in cryptocurrencies to users in return for taking back a stablecoin from them. 
However, unlike fiat collateral, the cryptocurrency price fluctuates, which can cause the discrepancy between the cryptocurrency price values that users and the system refer to due to non-zero transaction time and a discrete cryptocurrency price oracle update within the system. 
Thus, the value users earn can differ from 1.

Here, the cryptocurrency used as collateral is denoted by $c$.
Let $p^c_u$ and $p^c_s$ denote $c$'s price to which users and the system refer, respectively. 
In addition, the total value of cryptocurrency reserves is determined by the characteristics of the economy, which is a fundamental state $\theta$. 
Therefore, the total value of cryptocurrency reserves is denoted by a strictly increasing function $V^c(\theta)$ of $\theta$; naturally, the condition of greater asset value and higher asset growth is stronger fundamentals. 
Then $v$ is $p^c_u/ p^c_s$ in the case of $Q \leq V^c(\theta)$; users receive $1 /p^c_s$ cryptocurrencies by redeeming a stablecoin only when the crypto reserves are not depleted.

In fact, a ratio $p^c_u/p^c_s$ is for the fluctuation of $c$'s price; a greater drop in $c$'s price implies a smaller value of the ratio below 1. 
The price change of crypto collateral is affected by an economic situation that can be represented by the state of fundamentals, and also the stablecoin redemption action of users.
Note that in the redemption process, the system should pay the users, which implies that a part of the crypto collateral should be unlocked and flow into the market.
This can lower the cryptocurrency price by increasing its circulating supply and/or bringing other secondary market impacts. 
Given this, we will denote $p^c_u/p^c_s$ by a function $r^c\left(Q, \theta\right)$, where $r^c$ is strictly decreasing for the first input $Q$ and strictly increasing for the second input $\theta$.
That is, $p^c_u/p^c_s$ decreases as users redeem more stablecoins in a short period, while $p^c_u/p^c_s$ increases as the economic condition is better. 
As a result, $v$ is expressed as $r^c\left(Q, \theta\right)$ when $Q \leq V^c(\theta).$

On the other hand, if $Q> V^c(\theta)$, users can consider two cases where they receive $r^c\left(Q, \theta\right)$ or not. Then because the probability of users to get the reward is $V^c(\theta)/Q$ under the assumption that users are symmetric, the expected value would be $$r^c\left(Q, \theta\right)\cdot \frac{V^c(\theta)}{Q}=r^c\left(Q, \theta\right)\times \frac{V^c(\theta)}{Q} + 0 \times \left(1-\frac{V^c(\theta)}{Q}\right).$$
Therefore, in this case, the value of $v$ that users expect is $r^c\left(Q, \theta\right)\cdot V^c(\theta)/Q$. 

Additionally, note that $r^c$ and $V^c$ depend on the cryptocurrency $c$ used as collateral; therefore, even for the same inputs, the values of $r^c$ and $V^c$ can differ by which cryptocurrency the system uses. 
For example, consider a system that employs a robust exogenous cryptocurrency such as Bitcoin or Ethereum. 
How many of these cryptocurrencies the system will unlock to the market in the redemption process does not significantly influence their price because their value comes from other external sources. Therefore, in that case, $Q$'s impact on $r^c$ can be negligible. 
On the contrary, it can become significant (i.e., $\Delta r^c(Q,\theta)/\Delta Q$ would be larger) when $c$'s value comes from the system's usage (e.g., endogenous cryptocurrencies designed to back stablecoins). 

\subsection{Algorithmic stablecoins}
This category of stablecoins uses an endogenous cryptocurrency that the system can directly mint and burn to stabilize the stablecoin price. 
As the most representative example, UST falls into this category. 
An algorithmic stablecoin system sells and buys stablecoins to the market at the price $1$, where the payment is processed with its endogenous cryptocurrency (e.g., Luna in the Terra system). 
However, as mentioned when describing crypto-collateralized stablecoins, the value that users receive may be different from $1$. 
Then, similar to crypto-collateral stablecoins, $v$ is $r^c(Q, \theta)$.
A different point with crypto-collateralized stablecoins is that crypto-collateralized stablecoins have $v$ as $r^c\left(Q, \theta\right)$ only when the reserves are not exhausted, while $v$ always has this value in algorithmic stablecoins because they can always pay users by minting their endogenous cryptocurrencies. 

\subsection{Over-collateralized stablecoins} 
This type of stablecoins containing USDX and DAI\footnote{More specifically, the current version of DAI has a mixed mechanism of crypto collateralization using other stablecoins and over-collateralization.} is also popular. 
They require users to deposit crypto collateral with a value greater than $1$ in the system when minting a stablecoin. 
Here, we can consider the minted stablecoins as the user's debt. 
The stablecoin system returns the deposited collateral to its owner only when he/she redeems stablecoins. That is, in that case, only debtors can redeem their stablecoins and receive a value from the system. 
Alternatively, if a liquidation process for some specific collateral starts due to a drop in the collateral values, other users can redeem stablecoins (i.e., pay back stablecoin debts) on behalf of the debtor in return for taking the collateral.
Given this, $v$ should depend on whether users are stablecoin debtors or non-debtors.

First, let $D^L(\theta)$ denote the total quantity of stablecoins that users \textit{should redeem} to the system in the liquidation process at one point. 
If $D^L(\theta)$ is zero, it means that a liquidation process was not triggered for any collateral at the moment. 
Meanwhile, if $D^L(\theta)$ is equal to the circulating supply of the stablecoin, it suggests that a liquidation process was initiated for all collateral, encouraging users to redeem all stablecoins in the market. 
We also use the notation $D_{u_i}(\theta)$ to denote the stablecoin debt of user $u_i$ that did not enter a liquidation process. 
Here, $D_{u_i}(\theta)$ would be an increasing function of $\theta$; under a bad economic condition, more collateral of that user would be at risk and enter liquidation. 
If $D_{u_i}(\theta)$ is zero, user $u_i$ cannot redeem its stablecoins unless a liquidation process exists for some collateral. Note that the fact that $D_{u_i}(\theta)=0$ implies that user $u_i$ is a non-debtor or its deposited collateral is in the liquidation progress. According to the definition of notations $D^L(\theta)$ and $D_{u_i}(\theta)$, their relationship is as follows: $D^L(\theta)+\sum_{u_i}D_{u_i}(\theta)$ should be the same as the total circulating supply of stablecoins. Therefore, when $D^L(\theta)$ is the circulating supply of the stablecoin, $D_{u_i}(\theta)$ would be zero for any user $u_i$.

Lastly, we consider the notation $o(\theta)$: Crypto collateral to be paid per stablecoin from the system to users has a value of $o(\theta)$ from the system's perspective.\footnote{Strictly speaking, $o(\theta)$ would be different by each pair of collateral and debt, but we simplify it by standardizing the value over a pair because it does not affect our main theoretical analysis and result.} 
For example, if the collateral value deposited by user $u_i$ is 1.5 times the stablecoin debt based on the system's price oracle, then $o(\theta)$ would be $1.5.$
The value of $o(\theta)$ would be greater than $1$ unless the state of fundamentals $\theta$ is too small, because the system requires collateral worth more than $1$ when issuing a stablecoin debt to a user.
As a result, $v$ of user $u_i$ would be expressed as follows: $r^c\left(Q, \theta\right)\cdot o(\theta)$ if $0< D^L(\theta)$ or $0<D_{u_i}(\theta)$, otherwise zero. 
That is, it means that users can be paid by redeeming stablecoins to the system, only when there are liquidation processes or they are stablecoin debtors whose deposited collateral did not enter a liquidation process. 
\section{Equilibrium for Each Stablecoin Type}
\label{sec:eq}

In this section, we will analyze the equilibria and price instability for each type of stablecoin. 
To focus on the equilibria changing with $v$ and $v^\prime$ (i.e., equilibria varied by a stablecoin design), we consider that users have little incentive to keep holding coins (i.e., $i(x)\approx x$).
We also assume that values of the future status variables including the future fundamental state $\theta^\prime$, which $v^\prime$ depends on in stablecoins using cryptocurrencies as reserves, can approximate values of the current status variables such as the current fundamental state $\theta$ to simplify the equilibrium analysis by reducing the dimension of parameters. If we consider the future fundamental state as a variable $\theta^\prime$ independent of $\theta$, our results would be extended to two dimensions of fundamental states $[\theta, \theta^\prime]$.

\begin{figure*}
    \centering
    \includegraphics[width=0.9\textwidth]{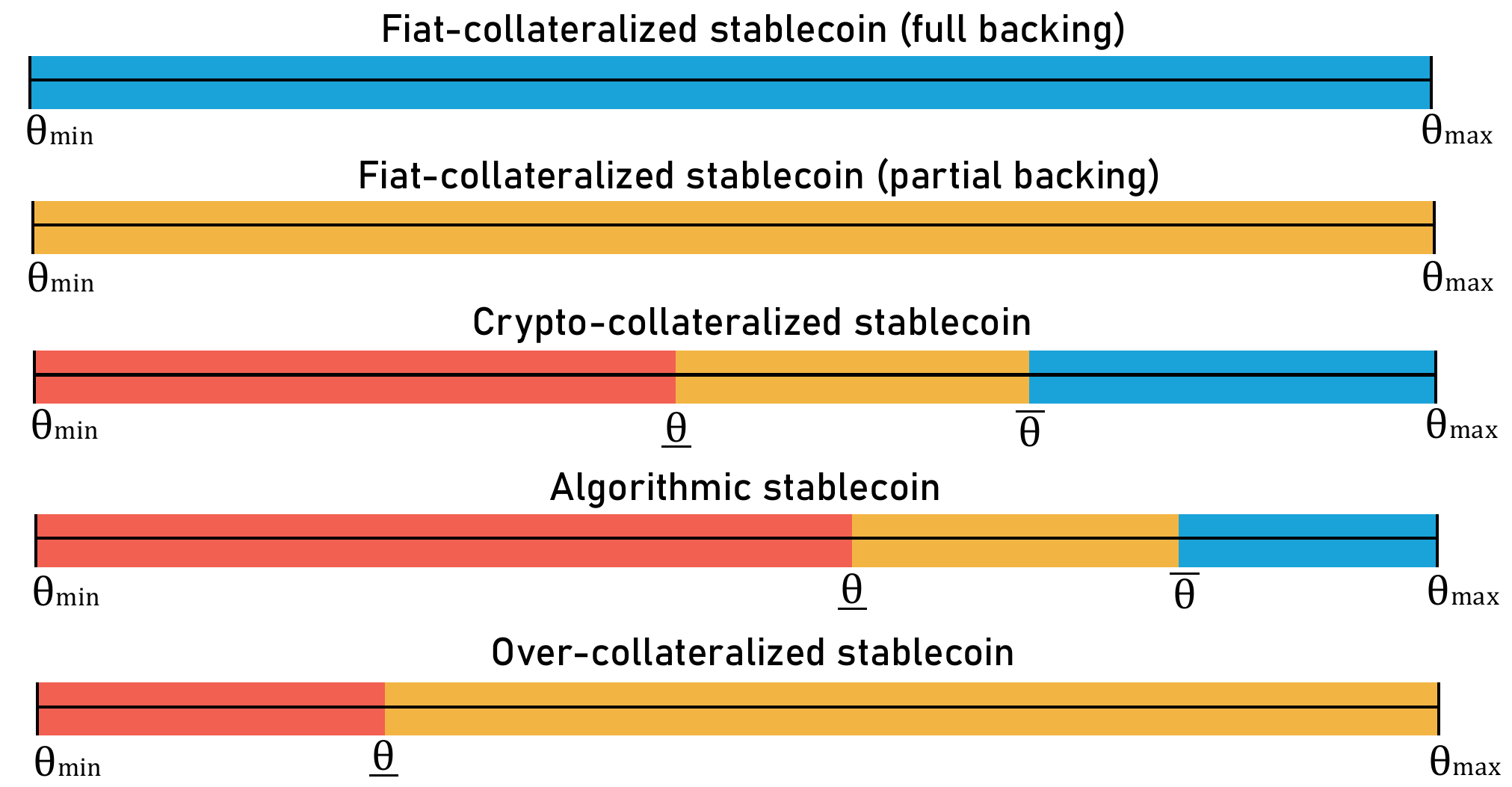}
    \caption{\textbf{An equilibrium state for given $\theta$ by stablecoin type.} 
    Each color bar represents a zone of $\theta$ with a different equilibrium state. In the blue zone, a unique pegging equilibrium exists. 
    In the yellow zone, there are multiple equilibria, including the pegging state of $p(M)=1$. In particular, the pegging state is a self-fulfilling equilibrium there.
    In the red zone, there is only a depegging equilibrium.}
    \label{fig:range}
\end{figure*}

We first present Figure~\ref{fig:range} that illustrates the result visually, which makes it possible to compare stablecoin designs intuitively.
In the figure, $\theta_{\max}$ and $\theta_{\min}$ indicate the maximum and minimum values of $\theta$, respectively. 
The blue bar represents a range of $\theta$ in which a unique pegging equilibrium exists so the peg is guaranteed. 
The yellow range of $\theta$ has multiple equilibria, including the pegging state of $p(M)=1$, which implies that the peg state is not guaranteed even though it can be reached. 
More specifically, in this range, the pegging state is a self-fulfilling equilibrium; users' belief totally determines the destiny of stablecoins.
Finally, in the red range of $\theta$, there is only a depegging equilibrium, which implies that the peg cannot be achieved. 
Therefore, the wider the blue range, the better the stablecoin design. 

A value of $\overline{\theta}$ indicates a lower bound for having a unique pegging equilibrium, and $\underline{\theta}$ is an upper bound for having only a depegging equilibrium. 
Therefore, according to Figure~\ref{fig:range}, stablecoins fully backed by fiat assets have $\overline{\theta}$ as $\theta_{\min}$. Stablecoins partially backed by fiat assets do not have a value of $\overline{\theta}$ and $\underline{\theta}$. 
Furthermore, an over-collateralized stablecoin does not have $\overline{\theta}$. 
In crypto-collateralized, algorithmic, and over-collateralized stablecoins, the values of $\overline{\theta}$ and $\underline{\theta}$ would depend on $c$. 
Specifically, if the system employs a more robust crypto asset in its pegging mechanism, the blue zone can widen by decreasing $\overline{\theta}$.
On a side note, a significant incentive function $i$ can also help expand the blue zone and narrow the yellow and red zones.

We now describe below the equilibrium state of stablecoin systems in detail.

\subsection{Fiat-collateralized stablecoins}

As described in Section~\ref{sec:model}, fiat-collateralized stablecoins have $v$ as $1$ if $Q\leq V^f,$ otherwise $V^f/Q.$ 
The following theorem presents equilibria of fiat-collateralized stablecoins. 

\begin{theorem}
Fully backed fiat-collateralized stablecoins have a unique pegging equilibrium for any $\theta$. On the other hand, for any $\theta$, partially backed fiat-collateralized stablecoins have multiple equilibria including the pegging state due to users' self-fulfilling beliefs.
\label{thm:fiat}
\end{theorem}

According to Theorem~\ref{thm:fiat}, fully backed fiat-collateralized stablecoins can guarantee the peg. 
Meanwhile, a system partially backing its stablecoin has multiple equilibria, so it is difficult to predict the consequence.
The system can reach the pegging state, but it is not guaranteed.
In particular, it suffers from a self-fulfilling belief of users; 
if users believe only a few stablecoin holders will redeem their coins, the users with that expectation would not need to redeem their coins to the system right now, which fulfills the expectation by themselves and, in turn, results in price stabilization. 
On the other hand, let us assume that users believe that too many holders will redeem coins so that the reserves of the system cannot cover it. 
Then users should immediately redeem their coins to the system, which realizes the expectation by themselves and puts the stablecoin in danger by bringing about the depreciation of the stablecoin.
We present the proof of Theorem~\ref{thm:fiat} in Appendix~\ref{app:proof2}. 

\subsection{Crypto-collateralized \& algorithmic stablecoins}

Next, we look at crypto-collateralized and algorithmic stablecoins.
Crypto-collateralized stablecoins have a value of $v$ as $r^c\left(Q, \theta\right)$ if $Q\leq V^c(\theta)$, otherwise $r^c\left(Q, \theta\right)\cdot V^c(\theta)/Q.$ 
In algorithmic stablecoins, $v$ is $r^c(Q, \theta)$.
We present their equilibria in Theorem~\ref{thm:crypto}. 

\begin{theorem}
We denote the total market supply of stablecoins by $T^s.$
Then, in crypto-collateralized and algorithmic stablecoins, $\overline{\theta}$ is a value such that $r^c\left(T^s,\overline{\theta}\right)=1$, and $\underline{\theta}$ is a value such that $r^c\left(0,\underline{\theta}\right)=1.$
Here, for crypto-collateralized stablecoins, we assume that $V^c(\underline{\theta})\geq T^s$.
Then both crypto-collateralized and algorithmic stablecoins have a unique pegging equilibrium for any $\theta\geq\overline{\theta}$, multiple equilibria including the pegging state for any $\theta$ in the range $[\underline{\theta},\overline{\theta})$, and depegging equilibria for any $\theta<\underline{\theta}$. 
\label{thm:crypto}
\end{theorem}

Crypto-collateralized and algorithmic stablecoins can guarantee the peg under good economic conditions. 
However, under poor economic conditions, even if the reserves are sufficient to back the stablecoins fully, they would not be able to reach the pegging state because users who redeem their coins cannot, in effect, receive 1 due to a downward price fluctuation of the cryptocurrency, the payment medium.
In the mediocre economic status, there are multiple equilibria including the pegging state; the successful peg is up to users' belief in others' redemption actions because the stablecoin redemption affects whether a cryptocurrency price can be less than 1 in that economic condition. 

The values of $\overline{\theta}$ and $\underline{\theta}$ depend on $r^c$ according to Theorem~\ref{thm:crypto}; $\overline{\theta}$ and $\underline{\theta}$ are the values such that $r^c\left(T^s,\overline{\theta}\right)=1$ and $r^c\left(0,\underline{\theta}\right)=1$, respectively. Note that $\overline{\theta}$ is greater than $\underline{\theta}$ because $r^c$ decreases and increases as the first and second inputs increase, respectively. 
The more robust cryptocurrency the systems use to back stablecoins, the gentler the slope of $r^c$. 
That is, for a more robust cryptocurrency $c$, $r^c$ can be greater, so $\overline{\theta}$ and $\underline{\theta}$ become lower. 
For example, crypto-collateralized stablecoins can use Bitcoin, Ethereum, or even other stablecoins as their collateral to widen the blue zone.
Meanwhile, algorithmic stablecoins should use endogenous cryptocurrencies according to their protocol, which would have a narrower blue zone and a wider red zone. 
We present the proof of the theorem in Appendix~\ref{app:proof3}.

\subsection{Over-collateralized stablecoins}

Finally, we consider over-collateralized stablecoins.
The stablecoin has $v$ for user $u_i$ as follows: $r^c\left(Q, \theta\right)\cdot o(\theta)$ if $0< D^L(\theta)$ or $0< D_{u_i}(\theta)$, otherwise zero. 
Here, $o(\theta)$ is greater than $1$ as long as $\theta$ is not too low. 
Theorem~\ref{thm:over} presents equilibria of over-collateralized stablecoins. 

\begin{theorem}
We assume that $r^c(0,\theta)\cdot o(\theta)<1$ when $D^L(\theta)=T^s$.
Then over-collateralized stablecoins have multiple equilibria including the pegging state for any $\theta\geq\underline{\theta}$, and depegging equilibria for any $\theta<\underline{\theta}$, where $\underline{\theta}$ satisfies $r^c(0,\underline{\theta})\cdot o(\underline{\theta})=1$. Moreover, $\underline{\theta}$ of over-collateralized stablecoins is smaller than that for crypto-collateralized and algorithmic stablecoins.
\label{thm:over}
\end{theorem}

In over-collateralized stablecoins, stablecoin debtors can redeem their coins anytime, while non-debtors cannot unless a liquidation process starts for some collateral. 
Therefore, users' belief regarding redemption by debtors plays an important role in maintaining the peg; if users believe that many stablecoin debtors will redeem their coins, debtors with this expectation will redeem their coins now to settle their debt cheaper because they expect an increase in the stablecoin price due to other debtors' redemption. This leads to a rise in a stablecoin price by fulfilling their expectation by themselves. 
On the contrary, if users believe that only a few debtors will redeem their coins, debtors with the expectation do not need to redeem their coins immediately because they do not think the coin price will increase. 
Therefore, the stablecoin price will not increase by vindicating their decision.

On the other hand, if the collateral value is less than 1 due to severely bad economic conditions, the system would not be able to attain the pegging state. 

Moreover, where a system pays users more than 1 based on its price oracle, $v$ can be not less than 1 even with a price drop of cryptocurrencies. This makes $\underline{\theta}$ of over-collateralized stablecoins smaller than that for crypto-collateralized and algorithmic stablecoins. 
We present the proof of Theorem~\ref{thm:over} in Appendix~\ref{app:proof4}. 
\section{Empirical Analysis}
\label{sec:data}

We found that existing stablecoin designs have different price equilibria by their unique value of $v.$ 
Here, we conduct an extensive empirical analysis using observational data to see whether actual stablecoin prices have fluctuated in agreement with our theory. To this end, we first compare the stability levels of stablecoins and examine the relationship between a stablecoin price and $v$. We publish the data in \cite{data}.

\subsection{Top 22 stablecoins}

In the analysis, we considered stablecoins that aim to be pegged at USD, and that existed for one year before the UST downfall (i.e., May 13, 2021, to May 12, 2022). 
We selected, based on market cap, the top 22 stablecoins among the ones satisfying the above criteria, and then collected their daily price data from CoinMarketCap~\cite{coinmarketcap}. 
Table~\ref{tab:coin} presents our analysis targets, their market cap, and some statistics of a daily price. 
In the table, the fiat-collateralized stablecoins are marked as \textit{Fiat}. \textit{Crypto-S} and \textit{Crypto-NS} indicate a crypto-collateralized system that uses other stablecoins and non-stablecoins as collateral, respectively. 
\textit{Algo} and \textit{Over} indicate an algorithmic and over-collateralized stablecoin, respectively. 
Some stablecoins combine different pegging mechanisms.
For example, DAI allows users to swap the coin with other stablecoins, such as USDC, in a 1:1 ratio, while having a loan mechanism to maintain the peg~\cite{dai_psm}.
LUSD also runs crypto-collateral and over-collateral mechanisms simultaneously~\cite{lusd_peg}. 

\begin{table}[!ht]
\centering
    \begin{tabular}{>{\centering\arraybackslash}m{0.13\textwidth}|>{\centering\arraybackslash}m{0.21\textwidth}|>{\centering\arraybackslash}m{0.12\textwidth}|>{\centering\arraybackslash}m{0.1\textwidth}|>{\centering\arraybackslash}m{0.1\textwidth}|>{\centering\arraybackslash}m{0.1\textwidth}} 
       Name & Type & Market Cap. & Avg. & Min & Max \\
         \hline
         \hline
         USDT & Fiat & \$67.56B & 1.0004 & 0.9959 & 1.0019 \\
         \hline
         USDC & Fiat & \$51.72B & 1.0000 & 0.9982 & 1.0016 \\
         \hline
         BUSD & Fiat & \$19.91B & 1.0001 & 0.9981 & 1.0037 \\
         \hline
         DAI & Crypto-S$\bm{+}$Over & \$6.87B & 1.0004 & 0.9878 & 1.0098 \\
         \hline
         TUSD & Fiat & \$1.62B & 1.0000 & 0.9985 & 1.0024\\
         \hline
         FRAX* & Crypto-S & \$1.48B & 1.0003 & 0.9871 & 1.0682  \\
         \hline
         USDP & Fiat & \$945.18M & 1.0001 & 0.9907 & 1.0060  \\
         \hline
         USTC** & Algo & \$637.41M & 0.9977 & 0.4086 & 1.0098  \\
         \hline
         USDN* & Crypto-NS & \$629.60M & 0.9866 & 0.7831 & 1.0157 \\
         \hline
         FEI & Crypto-S & \$422.23M & 0.9964 & 0.9468 & 1.0127  \\
         \hline
         GUSD & Fiat & \$363.84M  & 0.9972 & 0.9656 & 1.0319 \\
         \hline
         LUSD & Crypto-NS$\bm{+}$Over & \$174.97M & 1.0039 & 0.9515 & 1.0415 \\
         \hline
         HUSD & Fiat & \$160.19M & 1.0000 & 0.9976 & 1.0026  \\
         \hline
         USDX & Over & \$105.40M & 0.9676 & 0.6677 & 1.0203 \\
         \hline
         sUSD & Over & \$77.24M & 1.0002 & 0.9899 & 1.0276 \\
         \hline
         VAI & Over & \$54.92M & 0.8972 & 0.7408 & 1.0963 \\
         \hline
         CUSD & Crypto-NS & \$51.63M & 0.9991 & 0.9847 & 1.0230 \\
         \hline
         OUSD & Crypto-S & \$48.27M & 0.9985 & 0.9772 & 1.0481 \\
         \hline
         MUSD & Crypto-S & \$41.92M & 1.0024& 0.9111 & 1.0578  \\
         \hline
         RSV & Crypto-S & \$28.84M & 0.9996 & 0.9867 & 1.0196 \\
         \hline
         USDK & Fiat & \$28.82M & 1.0013 & 0.9795 & 1.0230 \\
         \hline
         EOSDT & Over & \$2.17M & 0.9625 & 0.4313 & 1.9084 \\
         \hline
    \end{tabular}
    \begin{tabular}{p{\textwidth}}
    \\\vspace*{-5mm}
    \renewcommand{\arraystretch}{1} 
    \linespread{1}\fontsize{9}{10.2}\selectfont
   *Although FRAX says that they are an algorithmic stablecoin, here we mark FRAX as Crypto-S, because it uses the algorithmic pegging mechanism only when the crypto-collateralized mechanism using other stablecoins does not work.\\
\linespread{1}\fontsize{9}{10.2}\selectfont
    **Here, USTC means UST before May 12, 2022. 
    \end{tabular}
    \vspace{2mm}
    \captionof{table}{Statistics of daily price data considering the time period from May 13, 2021, to May 12, 2022}
    \label{tab:coin}    
\end{table}

\subsection{Stability levels of stablecoins}
\label{subsec:stability}

Given that a time point corresponds to one value of $\theta$, we can compare the actual stability levels of stablecoins with the ones expected by our theory.
We evaluate the stability level of stablecoins by analyzing how close their price has been to \$1 (or how much their price has deviated from \$1).
We use two metrics to measure a (downward) price deviation from 1.
The metric to estimate a price deviation is defined as  $\sqrt{\frac{\sum_{i=0}^{N-1} (p_i-1)^2}{N}}$, where $p_i$ indicates one price data point in the data set. 
That is, it means a standard deviation of price from 1.
The metric to estimate downward price deviation is defined as $\sqrt{\frac{\sum_{i=0}^{N-1} (\min(p_i-1,0))^2}{N}}$, which implies a price deviation considering only when a price falls below 1.
For FRAX, unlike others, it aims to keep the price within the range of \$0.9933 to \$1.0033 rather than \$1.
Therefore, to calculate its price deviation and downward price deviation from the target range $[0.9933,1.0033]$, we apply $\sqrt{\frac{\sum_{i=0}^{N-1} \min_{x\in[0.9933,1.0033]}(p_i-x)^2}{N}}$ and $\sqrt{\frac{\sum_{i=0}^{N-1} (\min(p_i-0.9933,0))^2}{N}}$, respectively.
Moreover, we ran an independent t-test for all pairs of stablecoins to rank the stablecoins statistically.
Table~\ref{tab:price} in Appendix~\ref{app:tab} presents the results. 
Figures~\ref{fig:volatility} and \ref{fig:volatility2} show price deviation and downward price deviation from 1 by stablecoin type, respectively.

\begin{figure}[!ht]
    \centering
    \begin{subfigure}[h]{0.7\textwidth}
         \centering
          \includegraphics[width=\textwidth]{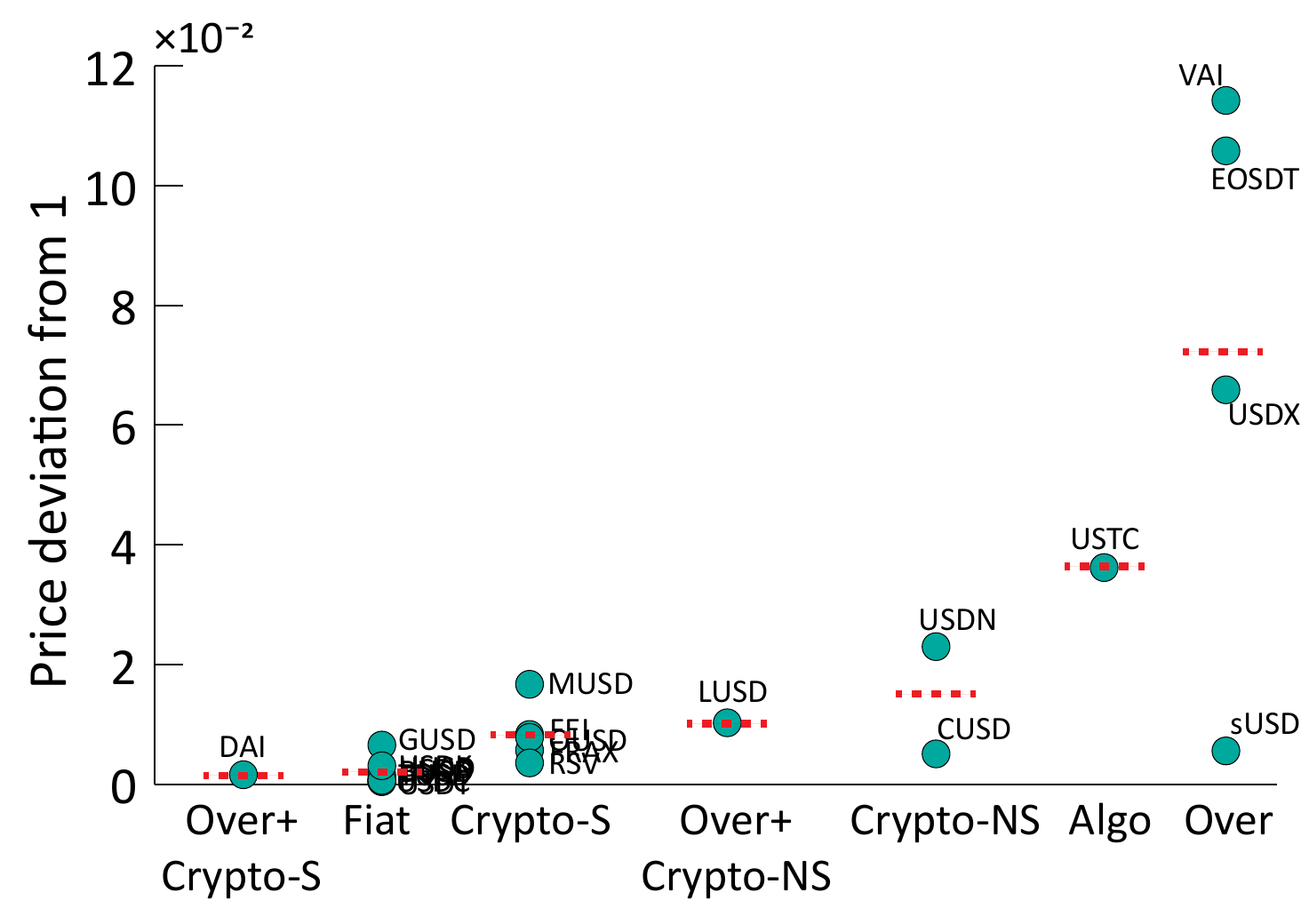}
         \caption{Price deviation from 1}
        \label{fig:volatility}
     \end{subfigure}\\
   \begin{subfigure}[h]{0.7\textwidth}
         \centering
          \includegraphics[width=\textwidth]{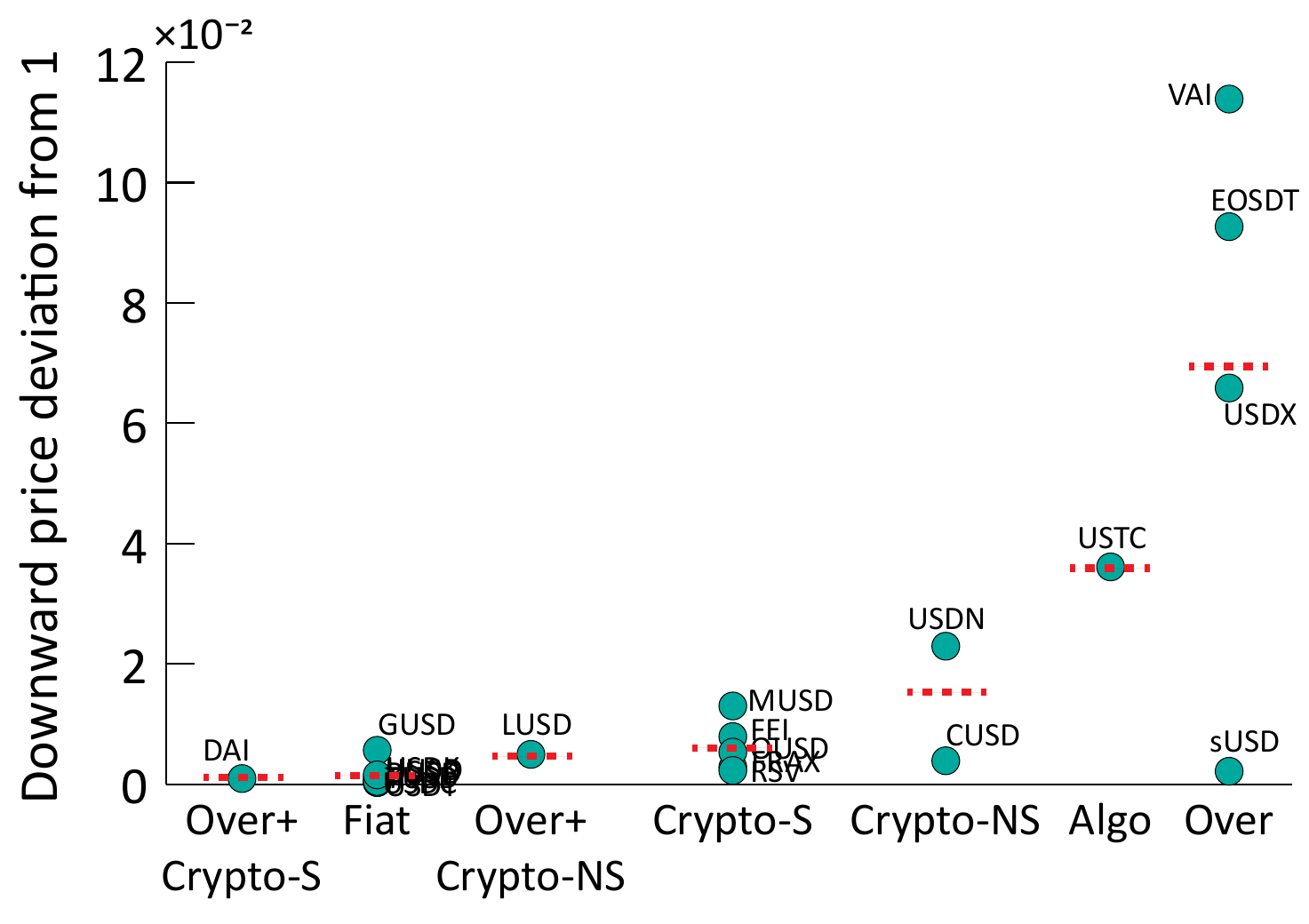}
        \centering \caption{Downward price deviation from 1 }
    \label{fig:volatility2}
     \end{subfigure}
     \caption{(Downward) price deviation by stablecoin type. The x-axis is arranged in ascending order of the average values by stablecoin type. Red dot lines represent an average value by stablecoin type.}
\end{figure}

Through the table and the figures, we can see that fiat-collateralized stablecoins (Fiat) and crypto-collateralized stablecoins (Crypto-S) using other stablecoins have been relatively stable. 
Over-collateralized stablecoins (Over) have been overall unstable, but there is also a large deviation across the samples. 
Note that the yellow zone that over-collateralized stablecoins possess largely in Figure~\ref{fig:range} has multiple equilibria, which makes it not easy to predict a price state.
On the other hand, because Crypto-NS and Algo categories have only one or two samples, it is difficult to compare them with other types meaningfully. 
However, we find that some algorithmic stablecoins have collapsed like UST or have changed their pegging mechanism; for example, Nubits have disappeared by not overcoming a severe peg break~\cite{nubits}. 
USDD also changed its original algorithmic pegging mechanism to the crypto-collateralized mechanism using combination of stablecoins and other cryptocurrencies~\cite{usdd,usdd2}.
Therefore, one may suspect that the general poor stability of algorithmic stablecoins results in the small sample number.

Another point we can see is that Crypto-S+Over and Crypto-NS+Over are observed to be more stable than Crypto-S and Crypto-NS, respectively.\footnote{Here, we would not claim the observation is statistically valid due to a small sample number.} In fact, this is also expected by our theoretical analysis. Theorem~\ref{thm:over} states that $\underline{\theta}$ of over-collateralized stablecoins is smaller than that for crypto-collateralized, which implies that crypto-collateralized stablecoins can reduce the red zone by combining with an over-collateralized mechanism. In addition, the combination of crypto-collateralized and over-collateralized systems does not affect the blue zone because all users can redeem their stablecoins in any case through the crypto-collateralized mechanism. As a result, the price stability of Crypto-S+Over and Crypto-NS+Over is empirically observed in agreement with our theory.

\subsection{Relationship between a price and $v$}

According to our theory, a stablecoin price should be depegged when $v$ is less than \$1, which should bring a correlation and causality between a price and $v$. 
We collected and explored actual redemption transactions of one year (05/13/2021$\sim$05/12/2022) to analyze the relationship between $v$ and a stablecoin price.
On-chain data collection was prioritized by high market cap, diversity of design, and ease of access to data.
In the process, we were able to collect on-chain data for a total of eleven stablecoins from five different blockchains, Ethereum, Terra, BSC, Celo, and Waves: DAI, FRAX, FEI, OUSD, MUSD, LUSD, USDN, CUSD, USTC\footnote{For USTC, we were able to collect redemption data for Oct 01, 2021 to May 12, 2022.}, sUSD, and VAI. 
The value of $v$ was calculated considering the quantity of the asset earned when users redeemed their stablecoins and the asset's market price. 
Because we were able to obtain only daily price data (i.e., end of day asset price) in USD, we used the last redemption transaction data for each day to reduce errors when calculating $v$.
On the other hand, for DAI and LUSD, we used the average value of $v$ for each day because they mix different pegging mechanisms. 
Here, note that we did not consider and subtract transaction fees by assuming that the difference between redemption and market transaction fees is negligible and by offsetting all the fees.\footnote{In fact, if we take into account transaction fees, the user payoff should be 1 minus transaction fees even when users sell a stablecoin in the exchange market.}
Therefore, for fiat-collateralized stablecoins, $v$ would be 1.

\begin{figure}[ht]
    \centering
    \includegraphics[width=0.6\columnwidth]{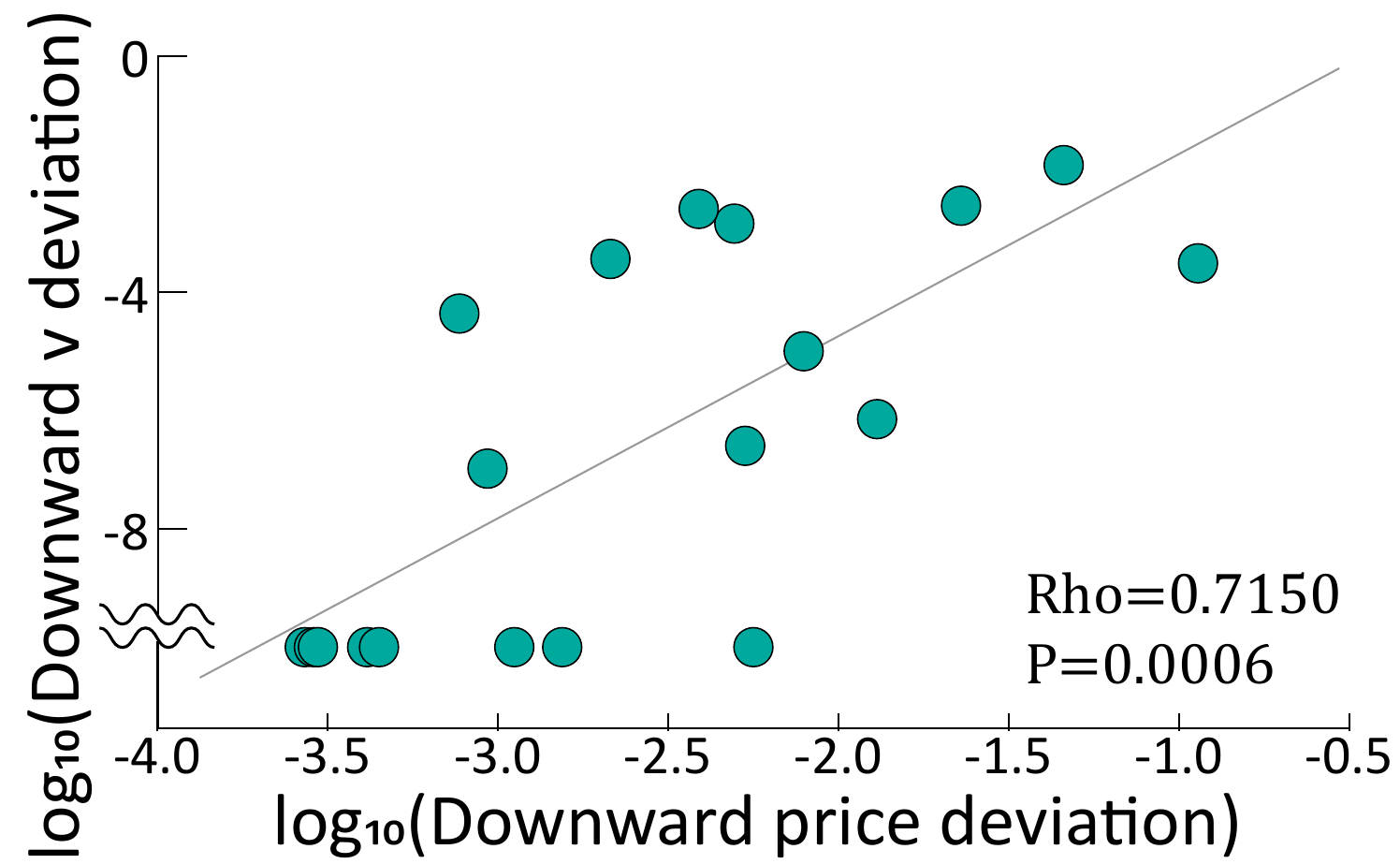}
    \caption{Correlation between downward deviations of $v$ and the price}
    \label{fig:corr}
\end{figure}

We first analyze the correlation between the downward $v$ deviation and the downward price deviation of stablecoins. 
Figure~\ref{fig:corr} shows that, considering 19 stablecoins (11 stablecoins, of which $v$ was actually collected, along with 8 fiat-collateralized stablecoins), there is a significantly strong correlation (Pearson's $\rho\approx$ 0.7150, p-value$\approx$0.0006, Bayes factor$\approx$0.0165\footnote{This value represents ``very strong evidence'' for the correlation~\cite{bayes}.}) between the downward $v$ deviation and the downward price deviation. 

\begin{table*}[ht]
    \centering
    \begin{tabular}{>{\centering\arraybackslash}m{0.12\textwidth}|>{\centering\arraybackslash}m{0.1\textwidth}||>{\centering\arraybackslash}m{0.1\textwidth}|>{\centering\arraybackslash}m{0.1\textwidth}||>{\centering\arraybackslash}m{0.1\textwidth}|>{\centering\arraybackslash}m{0.1\textwidth}}
        \multirow{2}{*}{\parbox{\linewidth}{\centering\vspace{1mm}Type} }&\multirow{2}{*}{\parbox{\linewidth}{\centering\vspace{1mm}Name}}& \multicolumn{2}{c||}{Correlation}& \multicolumn{2}{c}{Granger causality}\\
         \cline{3-6}
         & & Rho & P-value & F-stats.  & P-value\\
         \hline
         \hline
         Crypto-S$\bm{+}$Over&\cellcolor{YellowGreen}DAI & \cellcolor{YellowGreen}0.1499 & \cellcolor{YellowGreen}0.0136 & \cellcolor{YellowGreen}5.8753 & \cellcolor{YellowGreen}0.0160\\
         \hline\hline
          \multirow{4}{*}{\parbox{\linewidth}{\centering Crypto-S}}&FRAX & 0.1833 & 0.2576 & 1.1987 & 0.2809\\
          \hhline{~|*5{-}|}
         &\cellcolor{YellowGreen}FEI & \cellcolor{YellowGreen}0.1845 & \cellcolor{YellowGreen}0.0028 &  \cellcolor{YellowGreen}5.2799& \cellcolor{YellowGreen}0.0224\\
         \hhline{~|*5{-}|}
         &OUSD & 0.0934 & 0.2097 & 0.9743 & 0.3250 \\
         \hhline{~|*5{-}|}
         &MUSD & -0.0986 & 0.0620 & 1.2057 & 0.2729\\
         \hline
         \hline
         Crypto-NS$\bm{+}$Over &\cellcolor{YellowGreen}LUSD &\cellcolor{YellowGreen}0.3248& \cellcolor{YellowGreen}$<$0.0001 & \cellcolor{YellowGreen}30.1870 & \cellcolor{YellowGreen}$<$0.0001\\
         \hline
         \hline
          \multirow{2}{*}{\parbox{\linewidth}{\centering\vspace{0mm}Crypto-NS}}&\cellcolor{YellowGreen}USDN &  \cellcolor{YellowGreen}0.4914 & \cellcolor{YellowGreen}$<$0.0001 & \cellcolor{YellowGreen}76.9957 & \cellcolor{YellowGreen}$<$0.0001 \\
         \hhline{~|*5{-}|}
         &\cellcolor{YellowGreen}CUSD & \cellcolor{YellowGreen}0.1341 & \cellcolor{YellowGreen}0.0104 & \cellcolor{YellowGreen}12.3066 & \cellcolor{YellowGreen}0.0005  \\ 
         \hline
         \hline
         Algo &\cellcolor{YellowGreen}USTC & \cellcolor{YellowGreen} 0.8366 & \cellcolor{YellowGreen} $<$0.0001 & \cellcolor{YellowGreen}88.7618 & \cellcolor{YellowGreen}$<$0.0001\\
         \hline
         \hline
         \multirow{2}{*}{\parbox{\linewidth}{\centering Over}} &\cellcolor{YellowGreen}sUSD & \cellcolor{YellowGreen}0.7677 & \cellcolor{YellowGreen}$<$0.0001 &\cellcolor{YellowGreen} 44.1318 & \cellcolor{YellowGreen}$<$0.0001\\
         \hhline{~|*5{-}|}
         &VAI & -0.0200 & 0.7560 & 9.9877 & 0.0018\\
         \hline
    \end{tabular}
    \vspace{2mm}
    \caption{Correlation and Granger causality between a stablecoin price and $v$}
    \label{tab:v_correlation}
\end{table*}

In addition, for each stablecoin, we performed the correlation analysis between $v$ and a stablecoin price. 
Specifically, we examined the relationship between the last value of $v$\footnote{Remind that, for DAI and LUSD that combine different pegging mechanisms, we used the average value of $v$ for each day. } and the closing stablecoin price for each day, and used the minimum value between $v$ and 1 and the minimum value between a price and 1 to consider a downward fluctuation. 
We also saw if $v$ has affected the stablecoin price through the Granger causality analysis.
Table~\ref{tab:v_correlation} presents the results, where we colored coins if they have significant correlation and causality.

We can see that the significant correlation and causality between $v$ and a price are manifested in most stablecoins. 
DAI, FEI, LUSD, USDN, CUSD, USTC, and sUSD showed a significant correlation and causality. 
Meanwhile, it was not observed for FRAX, OUSD, MUSD, and VAI. 
In particular, we find that stablecoins with relatively good stability of a price and $v$ do not show a strong correlation and causality; 
overall, DAI and Crypto-S have a relatively high P-value in Granger causality.
This could be resulted from the deviations caused by noise or other factors being more apparent in coins with good downward stability of $v$.
Most representatively, we can think of fiat-collateralized stablecoins where $v$ is always 1. Definitely, their price deviations do not come from $v$. 

In addition, the existence of correlation and causality between $v$ and a price can be inconsistent across over-collateralized stablecoins; in our data, sUSD showed a significantly positive correlation and causality, but VAI showed an insignificant correlation. 
As described in Section~\ref{subsec:stability}, it can be difficult to predict consequences of over-collateralized stablecoins due to their wide yellow zone.

The last important point is that stablecoin systems with a popular and large incentive protocol have little correlation and causality between a price and $v$, which Theorem~\ref{thm:unique} points out.  
In fact, even though UST showed significant correlation and causality considering its collapse event, there was no significant correlation (Pearson's $\rho=$0.0421, P-value$=$0.5369) and causality (F$=$0.8426, P-value$=$0.3597) when only considering the period (Oct 01, 2021 to May 06, 2022) that an incentive protocol, Anchor, was greatly popular. 
Note that Anchor promises to give users nearly 20\% annual percentage yield (APY), and it even held about 75\% of the total UST market cap in some cases. 

We recognize the limitation of empirical analysis to show the true causality because there are not only two variables, $v$ and a price, in the real world. In fact, it is well known that deriving the true causality is really challenging.
By analyzing the stablecoin price along with many variables other than $v$, we will be able to examine whether the true causality between $v$ and the price exists.
Nevertheless, we believe that the results of the empirical analyses confirm our theory to some extent.
\section{Design Challenges}
\label{sec:discuss}

Our theoretical and empirical analyses indicate that, to improve their stability, stablecoin systems need to implement a robust incentive protocol, offer higher redemption payments when the asset to be used for a payment depreciates, or utilize more stable assets to back their coins. However, these measures may introduce other challenges.

Firstly, implementing a robust and popular incentive protocol can incentivize users to retain their stablecoins even during price declines. However, this approach can necessitate a substantial budget for providing sufficient monetary incentives to users. This issue has been highlighted in the case of Anchor, the popular incentive mechanism of Terra, where concerns about sustainability arose. Consequently, the design of incentive protocols becomes a crucial consideration. Developers may also explore non-monetary incentives, such as social reputation, as an alternative approach. However, it remains to be seen whether non-monetary incentives will be as effective as monetary incentives in this context.

Second, paying a higher value to users who redeem coins can reduce the difference between users' recognized value and the value intended by the system, even if the price of the asset, which is a payment medium, drops. This brings about a similar problem as the first solution because systems should possess more reserves to pay users with a higher value. 

Lastly, backing stablecoins with stable assets mitigates the impact of adverse economic conditions. Systems can achieve this by considering off-chain assets or other stablecoins, which is a common practice among many systems. However, this approach introduces certain challenges, such as centralization issues similar to those experienced by Tether, or reliance on other systems. Relying on other systems can potentially lead to a cascading risk. For instance, in March of this year, a bank run on Silicon Valley Bank (SVB) caused a ripple effect, leading to the depegging of several major stablecoins~\cite{svb}. Several stablecoins such as DAI depending on USDC have consequently been affected by a depeg of USDC that held a significant portion of its reserves in SVB. 

The challenges associated with stablecoin designs highlight the difficulty of completely eliminating risks. As a result, it becomes essential to focus on effectively managing risks and exploring optimal risk management strategies. By continuing to advance our understanding of stablecoin dynamics, we can work towards creating more robust and secure systems for the future.

\section{Concluding Remarks}
\label{sec:conclusion}

%
In this paper, we developed a common theory to characterize the stability properties of many stablecoins, considering reserve asset types and redemption mechanisms.
A continuous price drop of the assets backing stablecoins can reduce the actual (or recognized) payoff that users get by redeeming stablecoins, which can contribute to peg stress even in fully backed systems. 
Stablecoin systems collateralized by volatile cryptocurrencies can often suffer from this, leading to a higher price instability level compared to fiat-collateralized systems. 
To avoid the depegging state, such systems may need to rely on additional incentive mechanisms, such as large interest rates that can motivate many users to keep holding their coins.
As stablecoins receive newfound scrutiny, our model helps to improve understanding of design differences and establishes guidelines for future stablecoin design.

\newpage
\bibliography{references}

\appendix
\newpage

\section{Proofs}

\subsection{Proof of Theorem~\ref{thm:unique}}
\label{app:proof1}

First, we will show that the conditions that 
\begin{equation}
    \max\{v,i(v^\prime),i(p(M^\prime))\}>p(M) \text{ if } p(M)<1
    \label{eq:first}
\end{equation} and 
\begin{equation}
    \max\{v,i(v^\prime),i(p(M^\prime))\}\geq 1 \text{ if } p(M)=1
    \label{eq:second}
\end{equation} are sufficient to ensure the reachable and unique pegging equilibrium. 
We prove that, under this condition, the state $p(M)=1$ is an equilibrium. Consider user $u_i$ who decided not to sell its coins to the market. Then its payoff would be $v$ or $\max\{i(p(M^\prime)),i(v^\prime)\}.$ Due to Eq.~\eqref{eq:second}, both $v$ and $\max\{i(p(M^\prime)),i(v^\prime)\}$ are not smaller than $p(M)=1$.
Therefore, the rational choice of $u_i$ is not to change its action to selling its coins to the market. 
Moreover, considering users who decided to sell their coins to the market, even if they change their action, it would increase $p(M)$, but the maximum value of $p(M)$ is 1 according to our assumption described in Section~\ref{sec:model}. Therefore, $p(M)$ would also not change from $1$ in this case.
As a result, $p(M)=1$ is an equilibrium under the given conditions.

Next, we prove that $p(M)<1$ is not an equilibrium under the given conditions. Here, we consider user $u_j$ who decided to sell its coins to the market. Because of Eq.~\eqref{eq:first}, the user should change its action to redeeming coins to the system or to keeping holding coins to increase its payoff, which increases $p(M).$
Therefore, $p(M)<1$ cannot be an equilibrium, and $p(M)=1$ is a unique and reachable equilibrium. 

Moreover, we show that Eq.~\eqref{eq:first} is necessary to have the reachable, unique pegging equilibrium. 
We assume that there are some values of $p(M)$ ($<1$) such that $\max\{v,i(v^\prime),i(p(M^\prime))\}<p(M)$.
Here, we denote the value of $p(M)$ by $y.$
Then when $p(M)=y,$ it cannot reach $1$ because it is the optimal decision of users to sell coins in the market, which cannot decrease the market supply $M$ of the stablecoin. 
As a result, Eq.~\eqref{eq:first} is necessary to have a reachable and unique pegging equilibrium.
$\qed$

\subsection{Proof of Theorem~\ref{thm:fiat}}
\label{app:proof2}

When $T^s$ denotes the total supply of stablecoins, $Q$ is always less than or equal to $T^s.$
Then, in a fully backed fiat-collateralized system, $V^f$ is always greater than or equal to $Q$ because $V^f \geq  T^s$.
Therefore, according to Theorem~\ref{thm:unique}, the system has a unique equilibrium as a pegging state, because $v$ and $v^\prime$ that are 1 are always greater than $p(M)$ when $p(M)$ is less than 1. 

Meanwhile, a system partially backing its stablecoin (i.e., $V^f < T^s$) has multiple equilibria. 
If users believe that $Q$ will not be greater than $V^f$, they will act in the direction of inducing the state $p(M)=1$ because their rational action is not only redeeming but also holding coins. Note that in this case, users expect to earn $1$ even if they decide to keep their coins because $v^\prime$ would be expected to be $1.$ However, if they believe $Q$ will be greater than $V^f$, they will act in the direction of inducing the depegging state $p(M)<1$. In that case, their rational action is to redeem their coins to the system before $Q>V^f,$ which vindicates their beliefs by depleting reserves (i.e., $V^f=0$). This also leads to $v^\prime=0.$  
If $V^f/Q<e(\theta)$, the price $p(M)$ would reach $e(\theta)$. On the other hand, if $V^f/Q>e(\theta)$, users have a rational action as coin redemption, which decreases the value of $V^f/Q.$
As a result, the equilibrium price at that time would be $e(\theta).$
$\qed$

\subsection{Proof of Theorem~\ref{thm:crypto}}
\label{app:proof3}

To ensure the peg, a system needs to satisfy $r^c\left(Q, \theta\right)>p(M)$ for any $p(M)<1$ according to Theorem~\ref{thm:unique}. That is, $r^c\left(Q, \theta\right)\geq 1$ is needed.
Let $\overline{\theta}$ and  $\underline{\theta}$ be a value such that $r^c\left(T^s,\overline{\theta}\right)=1$ and $r^c\left(0,\underline{\theta}\right)=1$, respectively. 
For crypto-collateralized stablecoins, we assume that  $V^c(\theta)=T^s$ at $\theta=\theta^\circ$, where $\theta^\circ$ is less than $\overline{\theta}$ and $\underline{\theta}$ according to the assumption that $V^c(\underline{\theta})\geq T^s$.
For any $\theta\geq \overline{\theta},$ because $v$ and $v^\prime$ are always greater than $p(M)$ when $p(M)<1$, there exists a unique pegging equilibrium $p(M)=1$ according to Theorem~\ref{thm:unique}.  

On the other hand, in crypto-collateralized stablecoins, if $r^c\left(0, \theta\right)< 1,$ $v\,(=r^c(Q,\theta), \text{ or } r^c(Q,\theta) \cdot V^c(\theta)/Q \text{ in the case where } V^c(\theta)<Q)$ is less than $p(M)$ for some $p(M)<1$.
Therefore, for any $\theta<\underline{\theta}$, 
the optimal choice of users would be to sell their coins in the market when $p(M)=1$, which decreases the market price from 1. 
As a result, the pegging state $p(M)=1$ is not an equilibrium in that range where $\theta<\underline{\theta}$. 

Moreover, if $\theta<\underline{\theta}$, a depegging state becomes an equilibrium. 
In a crypto-collateralized stablecoin, for given $\theta\, (<\underline{\theta}),$ if $r^c(0,\theta)\leq e(\theta),$ the state $p(M)=e(\theta)$ would be an equilibrium because $v$ is always equal to or less than $e(\theta)$.
Moreover, for $\theta$ in the range $\theta^\circ \leq \theta < \underline{\theta}$, $v$ is always $r^c(Q,\theta)$. 
Then, within the given range of $\theta,$ if $r^c(0,\theta)> e(\theta),$ there is $y \,(\geq e(\theta))$ such that the state $p(M)=y$ becomes an equilibrium, where the value of $y$ depends on users' belief on $Q.$ For example, if users believe that a value of $Q$ will be $Q_1$ such that $r^c(Q_1,\theta)> e(\theta),$ the value of $y$ would be $r^c(Q_1,\theta).$ This is because users would not change their actions in the state $p(M)=r^c(Q_1,\theta)$, as the payoff for all actions is the same due to $p(M^\prime)=r^c(Q_1,\theta)$ according to the assumption that users' expected future can  approximate the current state. 
Alternatively, if users believe that $Q$ will have a value as $Q_2$ such that $r^c(Q_2,\theta)\leq e(\theta),$ the value of $y$ would be $e(\theta)$ because $v$ is not greater than $e(\theta).$ 

If $\theta< \theta^\circ$ and $r^c(0,\theta)> e(\theta),$ there are multiple depegging equilibria including  $p(M)=y \, (>e(\theta))$ and $p(M)=e(\theta)$, depending users' belief. 
If users believe that not many stablecoin holders will redeem their coins (i.e., $Q_1< V^c(\theta)$ and  $r^c(Q_1,\theta)> e(\theta)$), the users with the expectation do not need to redeem their coins right not, which induces the state $p(M)=r^c(Q_1,\theta).$
In contrast, if users believe that many holders try to redeem coins (i.e., $r^c(Q,\theta)\cdot V^c(\theta)/Q\leq e(\theta)$), the users' rational action is to redeem coins before $r^c(Q,\theta)\cdot V^c(\theta)/Q\leq e(\theta)$, which, in ends, leads to the state $p(M)=e(\theta)$. 

On the other hand, unlike crypto-collateralized stablecoins, we do not need to differentiate two ranges of $\theta$, 
$\theta^\circ \leq \theta < \underline{\theta}$ and $\theta < \theta^\circ$, to find depegging equlibria in algorithmic stablecoins for any $\theta<\underline{\theta}$ because $v$ is always $r^c(Q,\theta).$ Therefore, the algorithmic stablecoins also have multiple depegging equlibria according to the process above.

Lastly, in the range $\underline{\theta}\leq\theta<\overline{\theta}$, systems have multiple equilibria, including the pegging state, depending on users' belief, because whether $r^c(Q, \theta)$ is less than 1 depends on $Q.$  Specifically, if users believe $Q$ is small enough so that $r^c(Q, \theta)\geq 1,$ $p(M)=1$ would be an equilibrium. Otherwise, a depegging state would be an equilibrium.
Therefore, users' belief in whether only a few holders will redeem their coins will determine the system's future. $\qed$

\subsection{Proof of Theorem~\ref{thm:over}}
\label{app:proof4}

Let $\underline{\theta}$ be a value satisfying $r^c(0,\underline{\theta})\cdot o(\underline{\theta})=1$.
The notation $\theta^\circ$ denotes the maximum value of $\theta$ such that $D^L(\theta)=T^s$. 
Note that $\underline{\theta}>\theta^\circ$, according to the assumption that $r^c(0,\theta)\cdot o(\theta)<1$ if $D^L(\theta)=T^s$. 
In other words, $D^L(\theta)<T^s$ for any $\theta\geq\underline{\theta}$.

In that range of $\theta\geq\underline{\theta}$, there are multiple equilibria including the pegging state, by users' self-fulfilling belief in others' redemption actions, because whether $v\,(=r^c(Q,\theta)\cdot o(\theta) \text{ or } 0)$ is less than 1 depends on $Q$ and whether a user is a \textit{good} stablecoin debtor. Here, good debtors mean users whose collateral did not enter a liquidation process.
To describe this in more detail, we first introduce the notation $\theta^\star$ that indicates a value of $\theta$ such that $r^c(T^s,\theta)\cdot o(\theta)=1.$ If $\theta<\theta^\star,$ whether $r^c(Q,\theta)\cdot o(\theta)$ is less than 1 depends on $Q,$ which leads to multiple equilibria including the self-fulfilling pegging equilibrium. This can be proven similar to Theorem~\ref{thm:crypto}. 

On the other hand, if $\theta\geq \theta^\star,$ whether $v$ is $r^c(Q,\theta)\cdot o(\theta)$ or $0$ is up to whether a user is a good stablecoin debtor. And users would decide their rational behavior depending on their belief in good debtors' redemption. Good debtors should redeem (i.e., pay back) their stablecoin debts optimally considering a coin price. After they redeem all of their debts or their collateral triggers liquidation, their $v$'s value becomes 0, changing the status of whether they are a good debtor. This makes the debtors choose coin redemption only when the action can give them the maximum benefit. Consequently, they would redeem coins when the price is low since the payoff difference between $v$ and $p(M)$ is large in that case. 
Given this, if good debtors believe that other debtors will redeem their coins now, they expect the coin price will increase, which encourages the debtors to redeem their coins now. Therefore, this vindicates their belief and makes the state $p(M)=1$ an equilibrium. Otherwise, if they believe that other debtors will not redeem their debts now, their rational action would be not to redeem coins now, which fulfills their belief.  As a result, for $\theta\geq \theta^\star,$ there are multiple equilibria, including the self-fulfilling pegging equilibrium. 

The value of $r^c(0,\theta)\cdot o(\theta)$ is less than 1 for any $\theta<\underline{\theta}.$
Therefore, in that case, $v\,(=r^c\left(Q, \theta\right)\cdot o(\theta) \text{ or } 0)$ is always less than $1$.
As shown in the proof of Theorem~\ref{thm:crypto}, it implies that there cannot be a pegging equilibrium. Moreover, similar to the proof of Theorem~\ref{thm:crypto}, it can be proven that there are depegging equilibria in that range of $\theta.$

Next, we show that $\underline{\theta}$ of over-collateralized stablecoins is smaller than that for crypto-collateralized and algorithmic stablecoins. In crypto-collateralized and algorithmic stablecoins, $\underline{\theta}$ is a value such that $r^c(0,\underline{\theta})=1$ according to Theorem~\ref{thm:crypto}.
In over-collateralized stablecoins, $\underline{\theta}$ is a value satisfying $r^c(0,\underline{\theta})\cdot o(\underline{\theta})=1$.
Because a collateral value would not drop below 1 if the corresponding asset appreciates, $r^c(0,\theta)<1$ if $o(\theta)<1.$
Therefore, $r^c(0,\theta)<1$ if $r^c(0,\theta)\cdot o(\theta)<1.$
This implies that $\theta$ should be less than $\underline{\theta}$ of crypto-collateralized and algorithmic stablecoins if it is less than $\underline{\theta}$ of over-collateralized stablecoins. 
As a result, over-collateralized stablecoins have a smaller value of $\underline{\theta}$ than that for crypto-collateralized and algorithmic stablecoins.
$\qed$

\newpage


\section{Tables}
\label{app:tab}

\begin{table}[!ht]
\centering
\begin{tabular}{>{\centering\arraybackslash}m{0.11\textwidth}|>{\centering\arraybackslash}m{0.1\textwidth}||>{\centering\arraybackslash}m{0.15\textwidth}|>{\centering\arraybackslash}m{0.1\textwidth}||>{\centering\arraybackslash}m{0.15\textwidth}|>{\centering\arraybackslash}m{0.1\textwidth}}
        \multirow{2}{*}{\parbox{\linewidth}{\centering\vspace{1mm}Type} }&\multirow{2}{*}{\parbox{\linewidth}{\centering\vspace{1mm}Name}}& \multicolumn{2}{c||}{\makecell{Price deviation \\ from 1}}& \multicolumn{2}{c}{\makecell{Downward price \\ deviation from 1}}\\
         \cline{3-6}
         & & Value & Ranking & Value & Ranking\\
         \hline
         \hline
         \multirow{8}{*}{\parbox{\linewidth}{\centering Fiat}}&USDT & 6.1961$\times 10^{-4}$ & 3 & 2.7112$\times 10^{-4}$ & 1\\
         \cline{2-6}
         &USDC & 4.1747$\times 10^{-4}$ & 1 & 2.8842$\times 10^{-4}$ & 2\\
         \cline{2-6}
         &BUSD & 6.5501$\times 10^{-4}$ & 5 & 4.1341$\times10^{-4}$ & 4\\
         \cline{2-6}
         & TUSD & 4.8585$\times 10^{-4}$ &2 & 2.9634$\times 10^{-4}$ & 3 \\
         \cline{2-6}
         &USDP & 1.4753$\times 10^{-3}$ & 6 & 1.1181$\times 10^{-3}$ & 8 \\
         \cline{2-6}
         &GUSD & 6.5357$\times 10^{-3}$& 13 &5.6325$\times 10^{-3}$ & 15 \\
         \cline{2-6}
         &HUSD & 6.2961$\times10^{-4}$ & 4 & 4.4792$\times 10^{-4}$ & 5 \\
         \cline{2-6}
         &USDK & 3.0760$\times 10^{-3}$ & 8 & 1.5460$\times 10^{-3}$ & 9  \\
         \hline
         \hline
         Crypto-S$\bm{+}$Over &DAI & 1.5766$\times 10^{-3}$ & 7 & 9.3304$\times 10^{-4}$ & 7\\
         \hline
         \hline
        \multirow{5}{*}{\parbox{\linewidth}{\centering\vspace{2mm}Crypto-S}} &FRAX & 4.1901$\times10^{-3}$ & 10 & 7.7119$\times 10^{-4}$ & 6\\
         \cline{2-6}
         &FEI & 8.3035$\times10^{-3}$ & 15 & 7.9128$\times10^{-3}$ & 16\\
         \cline{2-6}
         &OUSD & 7.8163$\times 10^{-3}$ & 14 & 5.3287$\times 10^{-3}$ & 14\\
         \cline{2-6}
         &MUSD & 1.6690$\times 10^{-2}$ & 17 & 1.3002$\times 10^{-2}$&  17\\
         \cline{2-6}
         &RSV & 3.5518$\times 10^{-3}$ & 9 & 2.2750$\times 10^{-3}$ & 11 \\
         \hline
         \hline
          Crypto-NS$\bm{+}$Over & LUSD & 1.0255$\times 10^{-2}$ & 16 & 4.9520$\times 10^{-3}$& 13\\
         \hline
         \hline
          \multirow{2}{*}{\parbox{\linewidth}{\centering\vspace{0mm}Crypto-NS}}&USDN & 2.2982$\times 10^{-2}$ & 18 & 2.2933$\times 10^{-2}$ & 18\\
         \cline{2-6}
         &CUSD & 5.0588$\times 10^{-3}$ & 11 & 3.8918$\times 10^{-3}$  & 12 \\ 
         \hline
         \hline
         Algo &USTC &3.6189$\times10^{-2}$& 19 & 3.6120$\times10^{-2}$ & 19\\
         \hline
         \hline
         \multirow{4}{*}{\parbox{\linewidth}{\centering Over}}&USDX & 6.5899$\times 10^{-2}$ & 20 & 6.5859$\times 10^{-2}$ & 20\\
        \cline{2-6}
         &sUSD & 5.5589$\times 10^{-3}$ & 12 & 2.1421$\times 10^{-3}$ & 10\\
         \cline{2-6}
         &VAI & 1.1425$\times 10^{-1}$ & 22 & 1.1389$\times 10^{-1}$ & 22 \\
         \cline{2-6}
         &EOSDT & 1.0582$\times 10^{-1}$ & 21 & 9.2677$\times 10^{-2}$ & 21\\
         \hline
    \end{tabular}
    \begin{tabular}{p{\textwidth}}
    \\\vspace*{-5mm}
    \renewcommand{\arraystretch}{1} 
\linespread{1}\fontsize{9}{10.2}\selectfont
    *When running a t-test to rank a price deviation of stablecoins, there was no statistical significance (i.e., P-value $>$ 0.1) for the following 18 pairs out of 231 ($={}_{22}C_2$) pairs: (USDT,BUSD), (USDT, HUSD), (HUSD, BUSD), (USDP,DAI), (DAI, FRAX), (USDK,FRAX), (USDK, RSV), (RSV, FRAX), (FRAX, sUSD), (FRAX, CUSD), (CUSD, sUSD), (OUSD, USTC), (OUSD,FEI), (FEI, USTC), (LUSD, USTC), (MUSD, USTC), (USDN, USTC), (EOSDT, VAI)\\
\linespread{1}\fontsize{9}{10.2}\selectfont
    *When running a t-test to rank a downward price deviation of stablecoins, there was no statistical significance  (i.e., P-value $>$ 0.1) for the following 16 pairs out of 231 ($={}_{22}C_2$) pairs: (USDT,USDC), (USDT,TUSD),(USDC, TUSD),(BUSD, HUSD),(FRAX,DAI), 
    (DAI,USDP), (DAI, USDK), (USDP, USDK), (sUSD, RSV), (CUSD, LUSD), (LUSD,GUSD), (LUSD,OUSD), (OUSD, GUSD), (FEI, USTC), (MUSD, USTC), (USDN, USTC)
    \end{tabular}
    \vspace{2mm}
    \captionof{table}{Price volatility and downward price volatility of stablecoins given daily price data from May 13, 2021 to May 12, 2022.}
    \label{tab:price}    
\end{table}

\end{document}